# Correcting topographic artifacts in three-dimensional ToF-SIMS: A practical guide from data-only reconstruction to correlative AFM

Jacob M. Shusterman and Anton V. Ievlev[a)]

*Center for Nanophase Materials Sciences, Oak Ridge National Laboratory,*

*1 Bethel Valley Road, Oak Ridge, Tennessee 37831*

[a)]Email: ievlevav@ornl.gov

## ABSTRACT

Three-dimensional time-of-flight secondary ion mass spectrometry (3D ToF-SIMS) builds chemical volumes by stacking images acquired between sputter cycles. Standard reconstruction assumes a flat starting surface and a single sputter rate, so samples with topography, or with layers that sputter at different rates, yield distorted interface positions and depth scales. This article is a practical guide to correcting these artifacts. We group the available corrections by the information they need: the SIMS data alone, atomic force microscopy (AFM) topography before and after the profile, or AFM topography also at intermediate stages. All three are applied to a model Pt/$TaO_x$/Ta/Pt/Ti crossbar memristor device measured with in situ SIMS/AFM at ten stages, together with the data preparation they share. Each correction method is evaluated using the known device geometry and the measured surface evolution. The data-only correction requires no additional instrumentation and restores relative topography, but its depth scale depends on nominal layer thicknesses and remains unconstrained below the last calibrated interface. AFM before and after the profile fixes the start and end surfaces but assumes uniform sputtering in between, which fails next to the device, where the surroundings change material during the profile. Multi-stage AFM correction follows these changes and reproduces the structure of the device, at the cost of instrument time and repeated imaging of a partially sputtered surface. A decision flowchart and a checklist help readers choose, validate and report a correction.

# I. INTRODUCTION

Secondary ion mass spectrometry (SIMS) combines high-resolution mass spectrometry with sub-micrometer lateral and nanometer depth resolution, and is used to map the distribution of elements and molecules in surfaces, devices, and bulk materials. Recent examples include batteries[1-3], ferroelectrics,[4-6] memristors,[7-9] and perovskite photovoltaics.[10-12] In a dual-beam time-of-flight (ToF) instrument, a pulsed primary ion beam rasters the surface and each pixel yields a mass spectrum; a second, sputter beam removes material between analysis scans. Stacking the resulting images gives a 3D chemical volume.

Conventional stacking introduces geometric errors when it assigns measurements acquired from a nonplanar, evolving surface to flat planes. Standard reconstruction places every analysis scan in a flat plane and converts scan number to depth with one sputter rate. Both assumptions fail for samples with raised features, rough surfaces, patterned devices, or cells, and for samples containing materials that sputter at different rates (Figure *1*). Secondary ions from features at different heights are recorded in the same plane, so raised features appear buried and flat interfaces appear curved or stepped. Topography also changes local sputter and ionization yields and surface charging,[13] and differential sputtering can create topography that was not present at the start.[14] The result is a volume whose interfaces and layer thicknesses are wrong, even when the selected ion signals correctly identify the constituent layers.

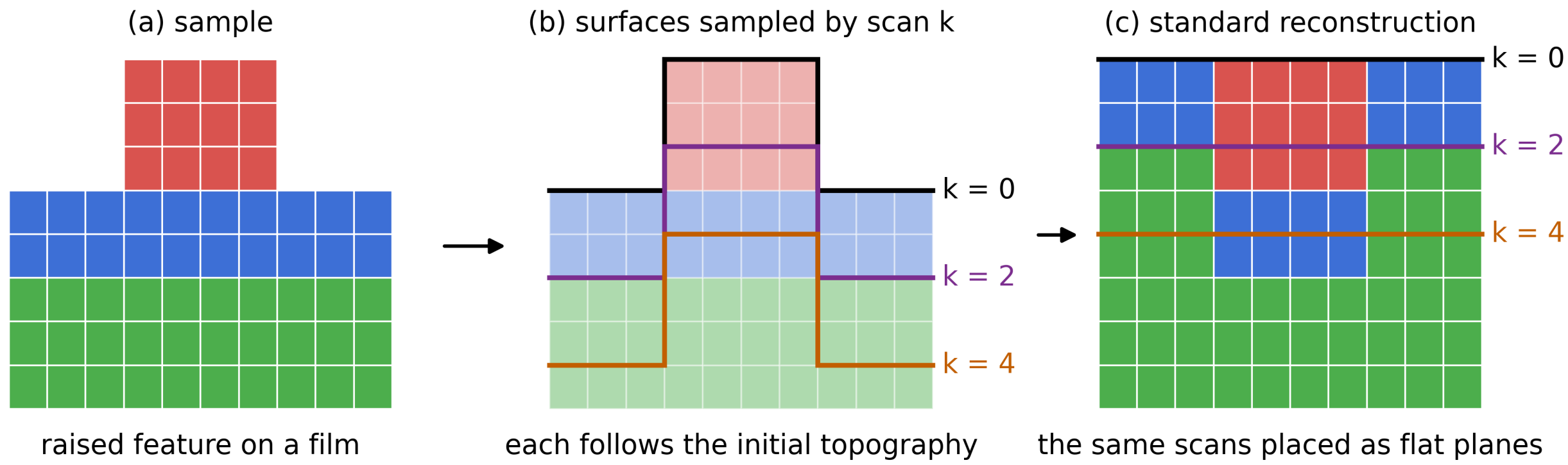

Figure 1. Origin of topographic distortion in 3D SIMS. *(a) Sample with a raised feature on a film. (b) The surface sampled by scan k lies k sputtered slabs below the initial surface in every column (uniform removal per cycle), so it follows the initial topography (lines for k = 0, 2 and 4). (c) Standard reconstruction places each of these surfaces in a flat plane: the raised feature appears level with its surroundings and the interfaces beneath it appear stepped. Colors represent different materials.*

Corrections for these artifacts have been developed in several communities. Correlative approaches add topography measured by AFM,[15-20] confocal microscopy (CM), or digital holographic microscopy (DHM),[21, 22] or reconstructed from helium-ion micrographs.[23] Data-only approaches use a buried reference interface, secondary electron (SE) images,[24, 25] total ion count (TIC) images,[26] or measured sputter rates.[27-30] Each paper typically describes one method on one sample type, and the methods differ in what they require, what they correct, and how they fail. For a new user facing a distorted 3D dataset it is therefore difficult to decide which correction applies, how to implement it, and how to tell whether it worked.

This article is intended to fill that gap. We compare substrate-referenced reconstruction, with and without nominal layer-thickness calibration, with single-stage and multi-stage AFM correction on one dataset; SE- and TIC-based methods are reviewed but not demonstrated. Section II classifies the sources of distortion, Section III organizes published corrections into tiers by the information they need (Table *1*) and gives a decision flowchart (Figure *2*). Sections IV – VIII then work through the implemented corrections on one dataset: a Pt/$TaO_x$/Ta/Pt/Ti crossbar memristor measured in a ToF-SIMS instrument with an AFM in the same vacuum chamber,[31] with topography recorded at ten stages of the profile. Section V describes the preparation steps that every tier requires, Sections VI – VIII the three tiers, and Section IX compares them quantitatively against information they do not use; finally, Section X provides some practical recommendations. Data and Python code that reproduce every figure and number are provided with the article.

## II. SOURCES OF TOPOGRAPHIC DISTORTION

It helps to separate four effects, because each correction addresses only some of them.

*Initial surface topography.* Height differences present before sputtering are the most common source of error. Standard reconstruction places the first scan of every pixel at the same height, so a feature of height $h$ appears displaced downward by $h$ relative to its surroundings, and every interface beneath it inherits the displacement (Figure *1*).

*Differential sputtering.* When materials in the analyzed volume sputter at different rates, a constant scan-to-depth conversion stretches slow-sputtering layers and compresses fast ones. Laterally heterogeneous samples also develop new topography during the profile, so the surface at scan $k$ no longer mirrors the initial surface.[14, 27] Sputter rates of the same material can also differ with its position in a stack,[27] and sputtering can be enhanced next to raised features, which digs a trench around them (Section V).

*Yield and charging effects.* Local surface slope changes sputter and ionization yields, and nonplanar insulating features can charge nonuniformly. Local geometry therefore affects both material removal and signal intensity; the intensity effects are not corrected by the geometric methods discussed here, but they complicate interface detection at sidewalls.[13]

*Lateral image shift.* Because the primary beam arrives at an oblique angle, the analyzed area moves laterally as the crater deepens. The shift is small for the 150–250 nm deep profile in this work but can become significant for deeper profiles, for example beyond about a micrometer, and it can be corrected in software or in hardware.[32] We treat it, together with the stage motion of profiles interrupted for AFM, as a registration step that precedes any topographic correction (Section V). Registration corrects rigid image translation only; oblique incidence also erodes the sidewalls of raised features asymmetrically (Section VIII).

The methods below correct the vertical (*z*) placement of voxels, that is, the first two effects. None fully corrects lateral distortions at steep sidewalls, which remain the main residual artifact in all corrected datasets shown here.

## III. OVERVIEW OF CORRECTION STRATEGIES

Published corrections differ mainly in what information they need beyond the SIMS data. We group them into four tiers (Table *1*). Higher tiers need more instrumentation and effort and replace some assumptions about how the depth evolves with measured surfaces, while registration, interpolation and extrapolation assumptions remain.

Table 1 Topographic correction strategies for 3D SIMS, ordered by the information they require. Tiers 0, 1a, 1a+, 2 and 3 are demonstrated in this work; Tier 1b is reviewed only.

| Tier / method | Required input | Key assumptions | What it corrects | Main limitations | Refs. |
|---|---|---|---|---|---|
| 0. Standard reconstruction with global depth scaling | SIMS data; crater depth or total thickness | Flat initial surface; one sputter rate | Global depth scale | No topography; distorts layers with different sputter rates | 13 |
| 1a. Substrate-referenced correction | Substrate (or buried reference) signal in every pixel | Reference interface is flat; vertical shift per pixel column | Initial topography (in scan units) | Fails without a reference signal, at low counts, and at sidewalls | This work |
| 1a+. Per-layer depth calibration | Interface positions; known layer thicknesses | Laterally uniform layers; constant rate within a layer | Scan-to-depth conversion; differential sputtering between layers | Needs prior thickness knowledge; cannot validate itself | This work,[27, 29] |

| Tier / method | Required input | Key assumptions | What it corrects | Main limitations | Refs. |
|---|---|---|---|---|---|
| 1b. SE- or TIC-based morphology | SE or TIC image at every cycle | Image contrast tracks local height | Topography at every cycle, including differential sputtering | SE images not always available and provide only approximate heights | [24-26] |
| 2. Single-stage correlative | Topography before and after the profile (AFM, CM, DHM, HIM); registration | Sputter front evolves linearly in each column between the two states | Initial topography; mean sputter rate per column | Misses nonlinear evolution and within-column rate changes; registration errors | This work, [15, 16, 20-23] |
| 3. Multi-stage correlative | Topography at intermediate depths, or an empirical sputter model | Sample survives intermediate imaging; stages can be co-registered | Layer-wise sputter rates; evolving topography | Time; tip-induced damage; cumulative registration error | This work, [17, 19] |

### A. Tier 0: standard reconstruction

Vendor software reconstructs the volume assuming a flat surface and converts scan number to depth using a single sputter rate, usually obtained from the final crater depth measured by profilometry or AFM, or from a known film thickness. This is adequate for samples with a sufficiently flat initial surface and an approximately constant sputter rate across the field and throughout the profile, and it should be the reference against which any correction is judged.

### B. Tier 1: data-only reconstruction

Tier 1 methods extract height information from the SIMS data themselves. Here, "data-only" means that no topography is measured: the per-layer variant (Tier 1a+) still uses nominal thicknesses, and for comparison with the other tiers the Tier 1 volumes are given one AFM height offset (Section VI-C). The substrate-referenced correction (Section VI) finds a buried interface assumed to be flat, typically the substrate, and shifts each pixel column so that the interface is leveled. Some instrument software offers a version of this; the manual implementation described here is useful when the substrate signal is noisy or when substrate ions also appear above the interface. If layer thicknesses are known, a per-layer depth calibration can follow. Alternatively, SE or TIC images recorded during the profile can be converted to height maps at every cycle.[24, 25]

SE images are usually recorded only in negative-ion mode, which is why the TIC variant was developed.[26] Tier 1 needs no additional instrument time, but every method in it relies on an assumption about the sample that cannot be verified from the SIMS data alone.

### C. Tier 2: single-stage correlative correction

Tier 2 methods measure the surface before and after the profile, co-register both topographies with the SIMS data, and distribute the voxels of each pixel column between the two surfaces (Section VII).[15, 16, 20] Among these methods, AFM gives the best lateral and height resolution and can measure deep craters, but scan areas are typically limited to about 100 × 100 $\mu m^2$. CM and DHM are noncontact and cover larger areas at sub-micrometer lateral resolution; DHM requires reflective or semi-reflective samples.[21, 22] Topography can also be reconstructed from helium-ion micrographs in instruments with in situ SIMS.[23] In principle, measurements can also be made ex situ,[15, 16] but relocating the same area after transfer is a major source of error, which hybrid instruments with AFM in the SIMS chamber reduce.[18, 31]

### D. Tier 3: multi-stage correlative correction

Tier 3 methods add topography measured at intermediate depths, for example after each layer of a multilayer stack, so that each segment of the profile is corrected separately.[17] An empirical sputter model can reduce the number of interruptions needed.[19] Tier 3 captures sputter-rate changes and evolving topography that Tier 2 averages out, at the cost of time and of repeated imaging of a partially sputtered surface (Section VIII).

Figure 2 turns Table 1 into a sequence of questions. After the data have been prepared and standard reconstruction has been ruled out, the next question is whether topography can be measured. Without topography, a flat reference interface allows the data-only correction (Tier 1a); otherwise only SE- or TIC-based methods remain. With topography, the contrast in sputter rates decides. For similar rates a single-stage correlative correction suffices, and Tier 1a is also adequate if a flat reference interface exists. Strongly different rates call for a multi-stage correction, provided the sample tolerates imaging during the profile; if it does not, single-stage correction remains an approximation whose errors must be checked against independent information (Section IX). A flat substrate alone therefore does not make topography measurements unnecessary, because Tier 1 cannot follow sputter-rate changes for which it has no thicknesses (Section VI).

Whatever the tier, three preparation steps come first: lateral registration of the image stack, removal of isobaric overlaps between the signals that represent the layers, and, for correlative tiers, placing the AFM data on an absolute height scale. They are described in detail in Section V, and errors made in them propagate into every corrected volume.

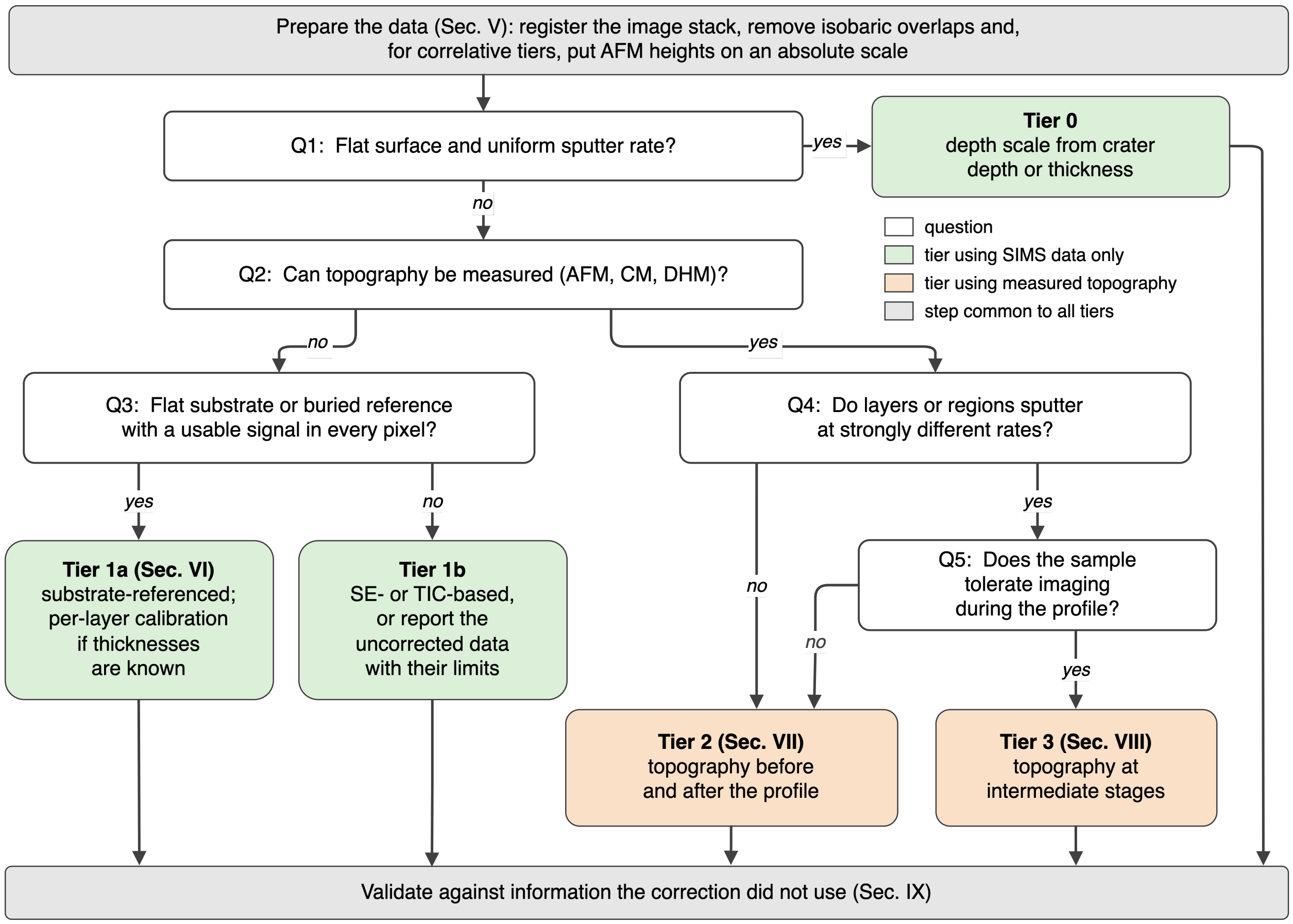

Figure 2. Decision flowchart for choosing a topographic correction. Tiers refer to Table 1. *White boxes are questions; green boxes are tiers that use the SIMS data only, orange boxes tiers that also need measured topography, and gray boxes steps common to all tiers. "Uniform" sputter rate means the same rate across the field and through the depth profile. Whatever path is taken,* the result should be validated against information not used in the correction (Section IX).

## IV. MODEL SYSTEM AND MEASUREMENTS

### A. Sample

All worked examples in this article use one measurement of one memristor device, so that every correction is applied to the same data and the data and code can be shared together.[33] The device is a pristine Pt/$TaO_x$/Ta/Pt crossbar memristor (Figure *3*). An 8 µm wide bottom-electrode (BE) line of 40-nm-thick Pt on a 10-nm-thick Ti buffer layer runs along *x*, a $TaO_x$/Ta film (30-nm-thick) covers the whole field, and an 8 µm wide top-electrode (TE) line of 60-nm-thick Pt runs along $y$. The full stack exists only in the $8 \times 8\ \mu m^2$ square where the two lines cross. The substrate is thermal $SiO_2$ on Si.

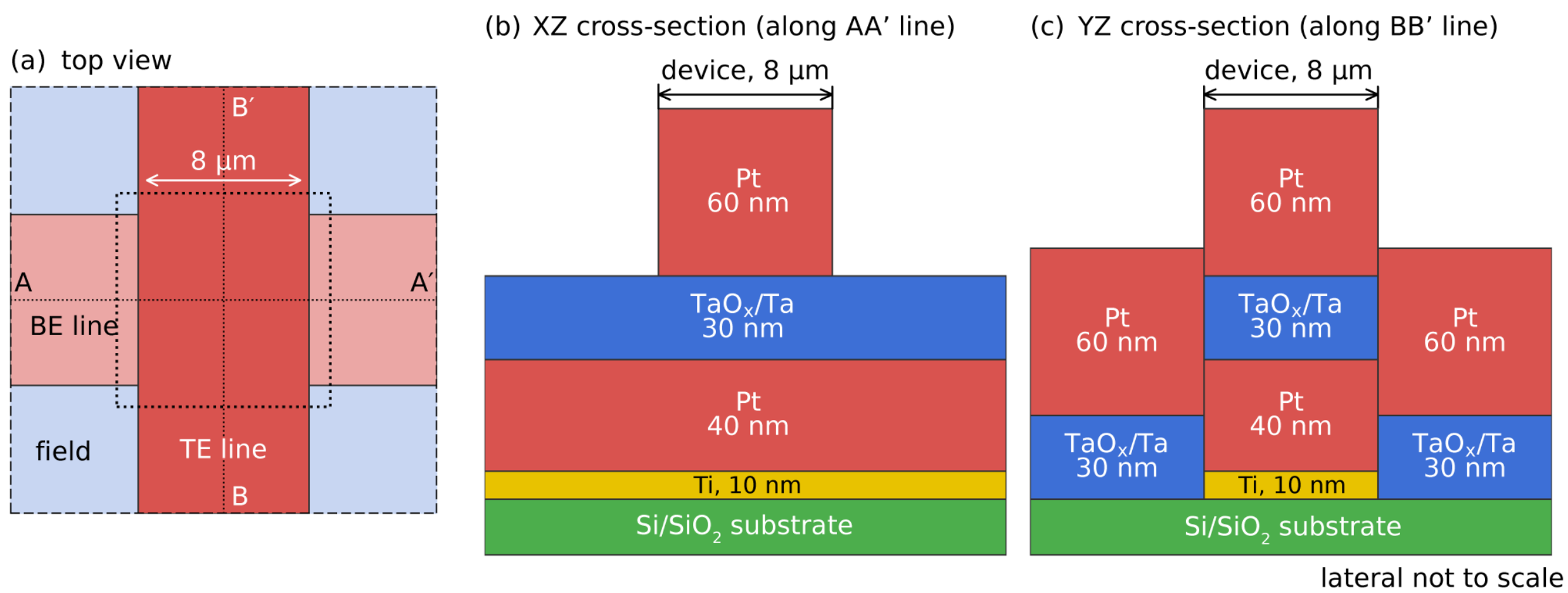

Figure 3. Model memristor device. (a) Top view of the crossbar: the BE line runs along x and the TE line along y, both 8 µm wide; the $TaO_x$/Ta film covers the field. Dashed and dotted squares mark the SIMS (20 µm) and AFM (10 µm) fields. (b), (c) Layer stacks along A–A′ (BE line) and B–B′ (TE line) *with nominal thicknesses: 60 nm top Pt, 30 nm $TaO_x$/Ta, 40 nm bottom Pt and 10 nm Ti on thermal $SiO_2$/Si.* Heights are to scale; lateral dimensions are not.

### B. ToF-SIMS

ToF-SIMS data were acquired with a TOF.SIMS 5-NSC instrument (ION-TOF GmbH, Germany) in positive-ion mode. A 30 keV $Bi_3^+$ beam was rastered over $20 \times 20\ \mu m^2$ with $256 \times 256$ pixels (78 nm per pixel), one shot per pixel and four frames per scan (19.7 s per scan). The analysis beam was operated unbunched (100 ns pulses), which gives high lateral resolution (~100 nm), but low mass resolution ($m/\Delta m \approx 80$–250; supplementary Fig. S1); the consequences for peak assignment are treated in Section V-B. Sputtering used a 1 keV $O_2^+$ beam (120 nA) rastered over $300 \times 300\ \mu m^2$ in non-interlaced mode, with 4 s of sputtering and a 0.5 s pause per cycle. A low-energy electron flood gun (21 eV) compensated surface charging. The profile comprises 240 analysis scans and 239 sputter cycles. Scans are numbered $k = 0 \ldots 239$, and scan $k$ is acquired after $k$ sputter cycles, so a position in the profile can be given as a scan index or as a cycle count, and a rate per scan equals a rate per cycle. Throughout, $k$ denotes the scan index, $K$ an interface position in scans ($K_s$ for the substrate interface), $t_m$ a layer thickness, $r_m$ a sputter rate in nm per cycle, $d$ a depth (positive downward) and $h$ a height (positive upward). Two features of the raw data matter below: $^{28}Si^+$ saturates the detector in the substrate (≈3 counts per voxel, i.e., ≈0.8 per shot; the measured $^{28}Si/^{30}Si$ ratio is 19 against the natural 30), so the substrate is followed using $^{30}Si^+$, and $Cs^+$ contributes about 10 % of all counts, a residue of $Cs^+$ sputtering of neighboring devices on the same chip.

### C. AFM measurements

Topography was measured using an AFM integrated in the same vacuum chamber as the SIMS. For each AFM measurement, SIMS acquisition was paused, the sample was moved from the SIMS analysis position to the AFM and back, and acquisition resumed. Topography was measured at ten stages: before sputtering; after 35, 70, 90, 110, 135, 157, 195 and 210 sputter cycles, chosen at or near layer interfaces; and after the last scan (239 cycles). The intermediate stages are assigned the cycle coordinates 35, 70, 90, 110, 135, 157, 195 and 210, i.e., the surface imaged at stage $c$ is taken to be the surface sampled by scan $c$. Because each interruption was made by hand, it may have fallen before or after sputter cycle $c$, which introduces a one-cycle uncertainty in this correspondence, 0.4–0.9 nm of depth, comparable to the residual scatter of the profile alignment (0.4–1.0 nm; Section V-C). At each stage two measurements were made: a 10 × 10 μm$^2$ image of the device (256 × 256 pixels, 39 nm per pixel), re-centered by hand on the device, and a 925 μm line profile across the whole crater, stitched from 14 linear AFM segments of ≈65 μm each. Images were recorded in contact mode with Multi75E-G probes (BudgetSensors; nominal spring constant 3 N/m) at a deflection setpoint of 200 nm, i.e., a nominal load of ≈600 nN. The images give local topography at high resolution; the profiles reach the unsputtered surface far outside the crater and therefore provide an absolute height reference at every stage (Section V-C). The device survived all eight intermediate measurements without visible tip damage; the roughness of the device top stayed at or below 2.4 nm.

The SIMS field (20 µm) is deliberately larger than the AFM field (10 µm). At a fixed number of pixels and dwell time, a larger SIMS raster costs little additional measurement time, keeps the device inside the field despite drift and stage motion (Section V-A), and provides registration features around the AFM footprint. The correlative corrections then use measured AFM heights inside the footprint and a model of the laterally uniform surroundings outside it (Section VII).

**D. Data processing and code**

Counts per channel were exported from the raw data to HDF5 files with the Python library Pyontof provided by the instrument vendor (ION-TOF GmbH), as integer arrays of 256 × 256 × 240 voxels without dead-time (Poisson) correction, together with the acquisition metadata. The correction would change the counts by less than 4 % in every channel used quantitatively here. The exceptions, the total ion count (up to 48 %) and $Cs^+$ (up to 75 %), enter only the edge statistic used for registration, where only the positions of edges are used. Registration on uncorrected counts assumes that dead-time losses, which are similar on the two edges of a line, do not shift the edge midpoints; this was not tested separately. These are whole-field factors; locally, in the substrate, $^{28}Si^+$ would need a much larger correction, which is why it is not used (Sec. V A). All further processing (registration, unmixing, the corrections and their validation, and every figure in this article) uses an open Python package written for this work and based on Python 3.12, NumPy 2.4, SciPy 1.17, h5py 3.16, Matplotlib 3.10 and pandas 3.0. Parameters that a reader may want to change, such as binning, smoothing, region definitions and reference regions for unmixing, are kept in one configuration file. Data and code are openly available.[33]

## V. DATA PREPARATION

Three preparation steps come before any correction. The image stack must be laterally registered; each layer must be represented by a signal free of isobaric overlaps; and, for correlative corrections, the AFM data must be placed on an absolute height scale. These steps do not depend on the correction tier, and errors made here can propagate into every corrected volume.

### A. Lateral registration

A profile interrupted for AFM contains three artifacts in the SIMS image stack (Figure *4*). (i) *Re-saturation transients.* In the first scan after each return, matrix-ion signals drop to 48–82 % of their stable level, while surface contaminants such as $K^+$ rise 1.9- to 5.1-fold (Figure *4a*); the $O_2^+$-saturated surface layer has to be re-established before the yields recover. By the next scan the matrix ions are back within ≈20 % of the stable level. (ii) *Motion jumps.* The sample does not return to exactly the same position, so the field shifts by up to 6 pixels (0.5 µm) in $x$ and 9 pixels (0.7 µm) in $y$ at the segment boundaries. (iii) *Thermal drift.* Within segments the field moves slowly, by up to 0.2 pixels per scan. The AFM can be re-centered on the device at every stage, but the SIMS field cannot be changed during a pause, so the device moves within it (Figure *4c,e*). Over the whole profile the accumulated shift is 2.0 µm in $x$ and ranges from −0.8 to +1.3 µm in $y$ (Figure *4b*).

Transient scans are flagged but kept. These scans remain part of the acquisition sequence and must retain their correct cycle coordinates; their intensities are not rescaled, which would be an analysis decision. The lateral shift of scan $k$ in segment $s$ is modeled as $J_s + \delta_s(k)$: a constant jump per segment ($J_s$) plus a smooth drift within it ($\delta_s$). Both are measured on the device edges. The edges are located with a multi-channel edge statistic, the sum over 13 channels [the total ion count, Pt (summed over its isotopes), $TaO^+$, $Ta^+$, $^{30}Si^+$, $SiO^+$, $^{48}Ti^+$, $TiO^+$, $Cs^+$, $K^+$, $Na^+$, $TaO_2^+$ and $TaH^+$] of the squared lateral derivative of the signal, each divided by its Poisson variance, which marks an edge whatever the contrast of the layer being sputtered (Figure *4c-f*). Raw intensities fail where the contrast inverts between layers. The $x$ shift follows the midpoint of the two TE-line edges and the $y$ shift the midpoint of the two BE-line edges. Midpoints are used rather than whole-image correlation because, as the sputter front crosses conformal layers on sloped sidewalls, the two edges of a line move apart or together; that is topography, not drift, and it cancels in the midpoint. Within each segment the per-scan midpoints are fitted with the polynomial of order 0–3 preferred by the Bayesian information criterion, and each jump is the step in the midpoint between the last stable scans of one segment and the first stable scans of the next, which are closest in depth and most similar in contrast. Transient scans are excluded from the estimation. Each scan is then shifted by an integer number of pixels, so counts are never interpolated, and the stack is cropped to the field common to all scans (230 × 229 pixels). Rounding the fitted shifts to whole pixels introduces a rounding residual of 0.3 pixels RMS, at most 0.5 pixels, in each axis; this does not include the uncertainty of the fitted shifts themselves.

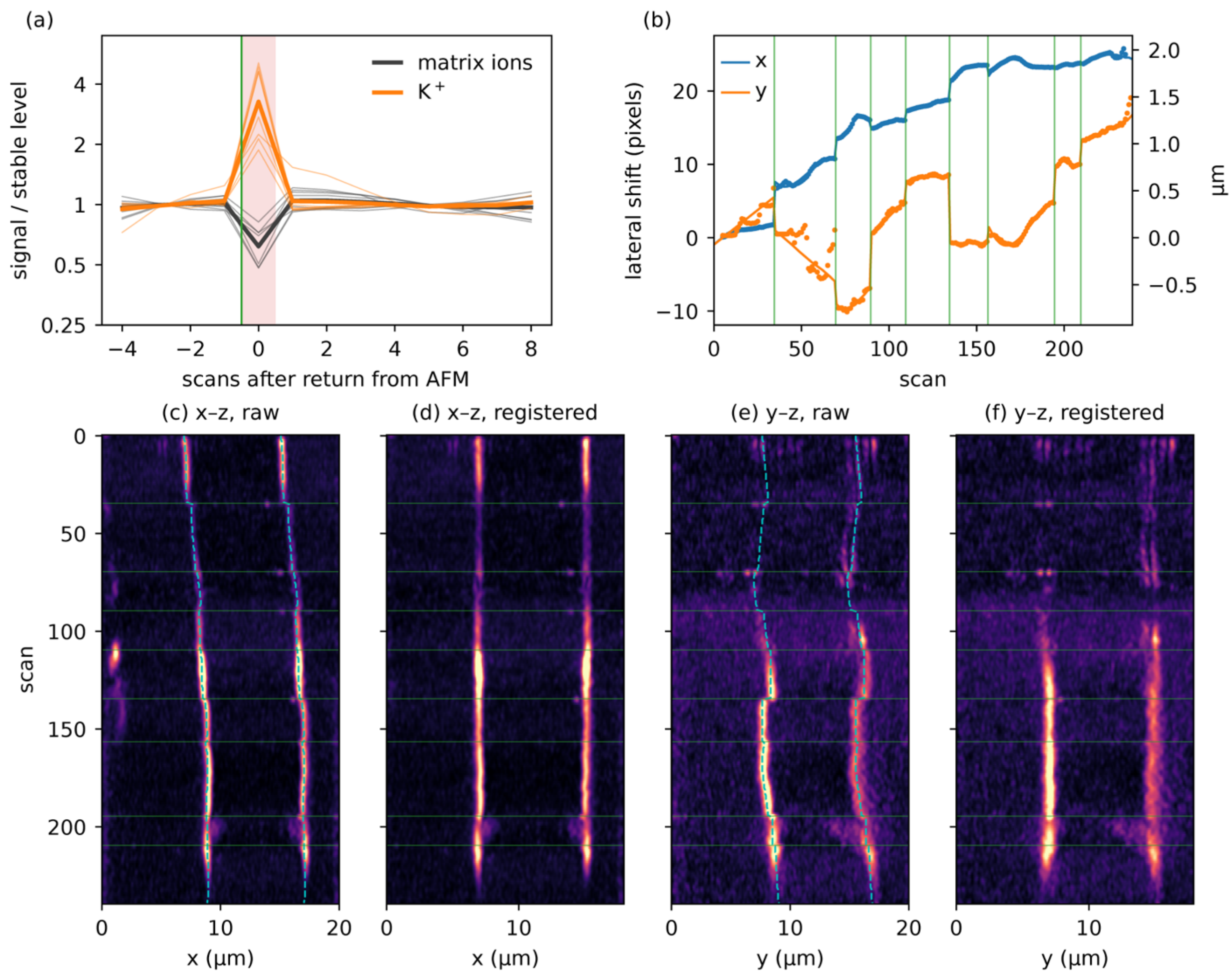

Figure 4. Artifacts of a depth profile interrupted for AFM, and their correction. (a) Matrix-ion signal (sum of $Pt^+$, $Ta^+$, $TaO^+$, $^{30}Si^+$ and $^{48}Ti^+$) and $K^+$ around the eight returns from the AFM, aligned at the first scan after the return (shaded) and divided by a stable level interpolated across the pause; thin lines are individual interruptions and thick lines their geometric mean. (b) Lateral shift of each scan in x and y: per-scan estimates (dots) and fitted model of segment jumps plus drift (lines); green lines mark the AFM interruptions. (c)–(f) The registration feature, a multi-channel edge statistic, in x–z sections through the device (TE-line edges, used for x) and y–z sections (BE-line edges, used for y), before (c), (e) and after (d), (f) registration; dashed lines are the edge positions predicted by the fitted model. The BE-line edges become visible only after scan ≈90.

The residuals of the fitted model are 0.03–0.31 pixels in $x$ and 0.2–0.6 pixels in $y$ for segments 3–9. In the first two segments (scans 0–69) the $y$ residuals reach ±4 pixels, because the edges of the BE line, the only $y$ contrast, are still buried (Figure *4e*); $y$ there is obtained by two-dimensional phase correlation of edge maps, using the chamfers and kinks of the TE line and particles. For the same reason this uncertainty barely affects the SIMS data in the region interiors; it affects the correlative corrections only within ≈0.3 µm of the device edges along *y*. The device measures 104 × 101 pixels, i.e., 8.1 × 7.9 $\mu m^2$ at the nominal pixel size, against 8.1 – 8.2 µm in the AFM images. Outline sizes depend on how the edges are defined and at which depth they are read, so they do not set the raster scale. The scale is fitted separately for *x* and *y* when AFM and SIMS are co-registered (Section VII-A). After registration, the edge trajectories are substantially aligned in both sections (Figure *4d,f*). The same procedure corrects the slower image shift of deep uninterrupted profiles.[32]

### B. Isobaric overlaps

At unit mass resolution, species with the same nominal mass cannot be separated by peak shape. In this sample, *m*/*z* 194–196 contain Ta adduct ions in addition to $Pt^+$, *m*/*z* 197 contains $^{196}PtH^+$ in addition to $TaO^+$, and *m*/*z* 198 contains $TaOH^+$ in addition to $^{198}Pt^+$. Isotope ratios reveal the overlap. In the top Pt layer, the counts at *m*/*z* 194, 195 and 196 are in the ratio 0.34 : 0.37 : 0.29, close to natural Pt (0.36 : 0.37 : 0.27, normalized over the same three isotopes). In the $TaO_x$/Ta film of the field, which has no Pt underneath, the same channels still carry about 13 000 counts, in the ratio 0.13 : 0.53 : 0.34. Displayed as Pt, this signal places Pt inside the oxide layer.

If each material produces a fixed pattern of counts across a set of mass channels, the counts $c$ of a voxel are a linear combination $c = Ea$ of the material patterns (the columns of $E$) with abundances $a$, as in multivariate analysis of ToF-SIMS data. We measured the patterns of three materials in regions of known composition: Pt in the top Pt of the device, $TaO_x$ in the field film near the surface, and Ta in the Ta layer of the device, using eight channels ($^{194–196}Pt^+$, $TaO^+$, $TaOH^+$, $Ta^+$, $TaO_2^+$ and $^{195}PtO^+$). Abundances were obtained by unconstrained linear least squares. A non-negativity constraint would look more natural per voxel, but at a few counts per voxel it cannot cancel noise and therefore biases the estimates upward; here it assigns 6–9 % of the signal to components that are absent. Under the fixed-pattern linear model, the linear estimate is unbiased once averaged over slices or regions, and is clipped to non-negative values only for display. Regions that were not used to define the patterns provide a check: after unmixing, the normalized Pt signal fraction, a signal estimate rather than a composition, is −0.4 % in the device $TaO_x$/Ta layer, 100.5 % and 101 % in the bottom Pt and the TE line, and 0.6 % in the $TaO_x$/Ta film on the BE line. $TaO_x$ and Ta are combined into one $TaO_x$/Ta layer in all figures, and Ti and the substrate are followed with $^{48}Ti^+$ and $^{30}Si^+$. The patterns and checks are listed in supplementary *Table S1*.

### C. Absolute AFM heights

AFM images give relative heights only: each image has its own offset and tilt, and a 10 µm image lies entirely inside the 300 µm crater. The long profiles provide the missing reference. Each profile is resampled onto the positions of the first (before-sputtering) profile, and a straight line fitted to the height difference over the unsputtered surface far outside the crater (0–230 and 700–925 µm) is subtracted; the residual scatter there is 0.4–1.0 nm. Heights are then expressed relative to that unsputtered surface. The device-top height read from the profile at each stage fixes the offset of the corresponding image, after the image has been registered to the first stage on its outline (shifts ≤3 pixels, 0.12 µm) and leveled on the device top.

Profiles and images measure the step between the device and the TE line independently, and agree within ≈1 nm up to the $TaO_x$ stage (50.8 and 50.1 nm before sputtering). From the Ta stage onward, the TE line read on the profile 20–40 µm from the device is up to 20 nm higher than the TE line read in the image within 1 µm of the device. The difference is a trench: next to the device the line is sputtered faster than far from it. Along the profile the trench is ≈20 nm deep within 1 µm of the device edge, ≈12 nm at 1–4 µm and ≈5 nm at 4–8 µm, and vanishes beyond ≈8 µm (supplementary Fig. S2). Because most of the 6 μm SIMS margin around the device lies in this zone, the trench is included explicitly when the surface outside the AFM footprint is modeled (Section VII).

## VI. TIER 1 IN PRACTICE: SUBSTRATE-REFERENCED CORRECTION

The substrate-referenced correction assumes that the substrate surface is flat, so any apparent step in the substrate interface is an artifact of the topography above it. Leveling that interface in every column restores the relative heights of everything above it. It needs only the SIMS data and, for the depth scale, nominal layer thicknesses. In the worked example the device column reaches the substrate last and the field first (Figure *5*).

### A. Step 1: locating the substrate interface

For each column (*x*, *y*), the interface is placed at the scan where the substrate signal rises most steeply, a criterion that locates interfaces accurately when counts are adequate:[26]

$$K_s(x,y) = \underset{k}{argmax}\, \frac{\partial S(x,y,k)}{\partial k}, \qquad (1)$$

where *S* is the $^{30}Si^{+}$ signal and *k* the scan index. Counts are the main obstacle. With unbunched acquisition the averaged substrate signal is below 0.1 counts per voxel, so a single-pixel profile is a sparse sequence of zeros and ones. Binning 4 × 4 laterally and 2 in depth, followed by Gaussian smoothing over σ = 2 bins, gives usable profiles. Even so, Eq. (1) fails wherever the rise is gradual. Outside the device the $SiO_2$/Si substrate gives a slow, two-step rise of $^{30}Si^{+}$, and the steepest point of a noisy profile can fall anywhere on it: the derivative map is visibly noisy (Figure *5b*). Figure 5a shows the binned and smoothed profiles of one typical bin in the device and on the TE line, at the positions marked in Figure 5b. At the sharp interface of the device both criteria agree; on the TE line the steepest point falls on a noise fluctuation after the rise. Over the bins of the TE line, BE line and field, the steepest-rise positions spread over 55–90 scans (interquartile range) and are biased toward late scans. At the available counts, the steepest-rise estimate is unstable in the broad substrate transition; with enough counts, for example on profiles averaged over a region, the two criteria agree within a few scans wherever the rise is sharp.

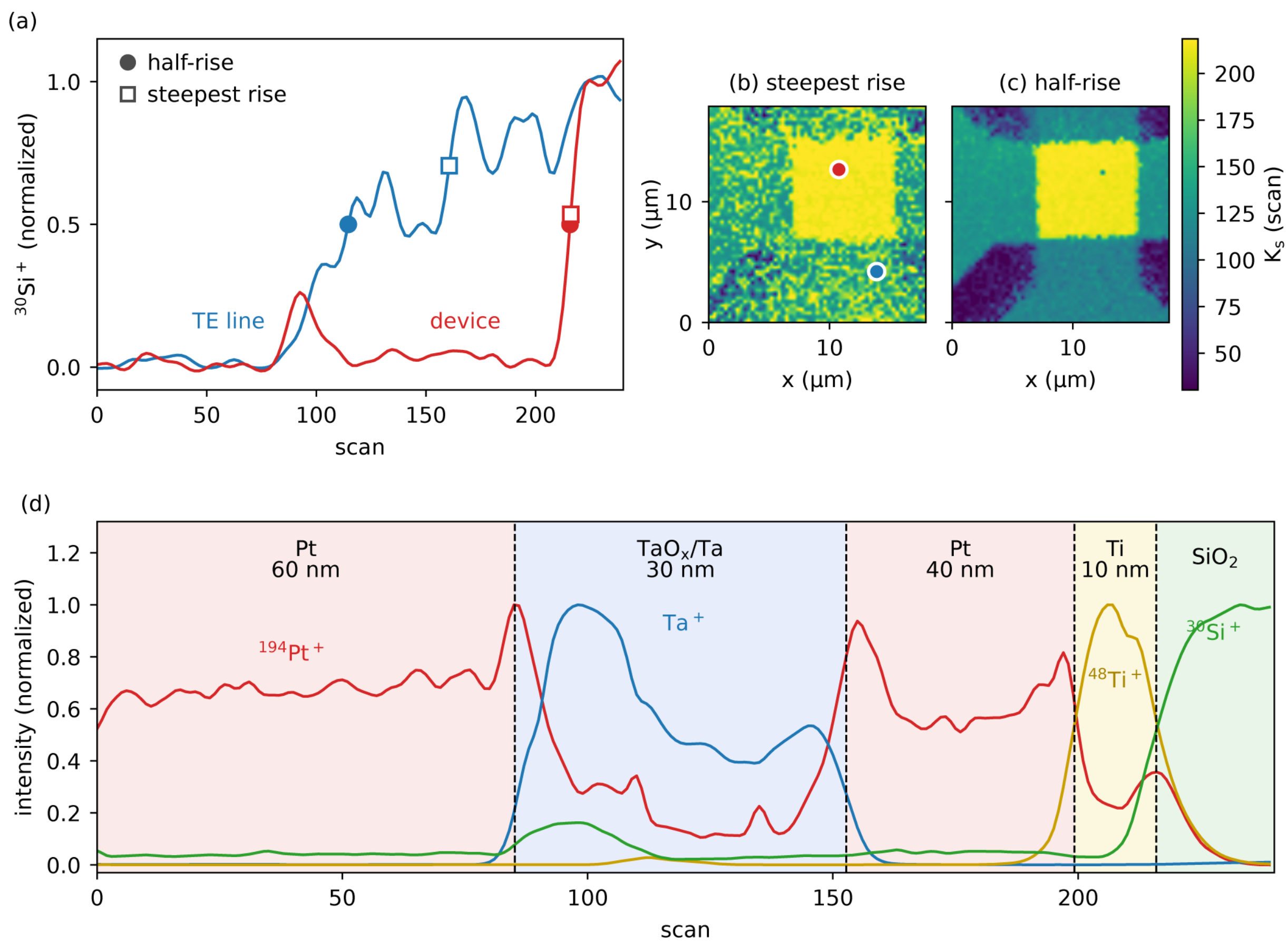

Figure 5. Tier 1 topography correction on the memristor device. (a) $^{30}Si^+$ depth profiles of one typical bin (4 × 4 × 2 voxels, smoothed over σ = 2 bins) on the TE line and in the device, normalized, with the substrate interface from the half-rise (filled circles) and steepest-rise (open squares) criteria; the two bins are marked in (b). (b), (c) Substrate-interface maps $K_s$ from the steepest-rise criterion of Eq. (1) and from the half-rise criterion, after binning 4 × 4 × 2 and smoothing over σ = 2 bins. (d) Normalized depth profiles of the device region with the four interfaces used for the per-layer calibration (dashed) and the nominal layers (shaded). Pt is shown as $^{194}Pt^+$, the isotope least affected by Ta adducts.

A more robust criterion is the first crossing of the midpoint between the baseline before the rise and the plateau after it ("half-rise"). In each binned profile, smoothed along depth only, the baseline is the median of the first 5 bins (10 scans) of the search window, which starts after the initial surface transient (scan 5), and the plateau is the median of the last 10 bins (20 scans); the interface is the first upward crossing of the midpoint, interpolated linearly between bins, and a profile that never crosses is assigned the end of the window. Binning and smoothing were set to the smallest values that give a stable map (supplementary Figure S3). It averages over the whole rise and, for a symmetric interface with adequate counts, gives the same position as Eq. (1). It places the interface at scans 215, 116, 130 and 53 (medians over each region) in the device, TE line, BE line and field, with interquartile ranges over the bins of a region of only 3–8 scans (17 scans in the field), and gives a clean map (Figure *5c*). As a rule, binning and smoothing should be increased only until the map is stable, and the map rather than individual profiles should be inspected, because larger kernels round the edges and remove features (supplementary Figure S3).

### B. Step 2: shift map and leveling

The shift map is the interface position relative to a reference region that should not move:

$$\Delta K(x,y) = K_s(x,y) - K_{ref}, \tag{2}$$

where $K_{\mathrm{ref}}$ is the median of $K_{\mathrm{s}}$ in the device. Every voxel is then moved to

$$k_{corr}(x,y,k) = k - \Delta K(x,y), \tag{3}$$

so that the substrate interface lies at $K_{\mathrm{ref}}$ in every column; columns that reached the substrate sooner ($\Delta K < 0$) move down. The shift map is −100, −85 and −162 scans in the TE line, BE line and field (medians over the same regions, computed before the interface positions are rounded). Without further information, depth is converted to nm with one rate for the whole volume, here the nominal device stack (140 nm, Section VI-C) divided by $K_{\mathrm{ref}}$, i.e., 0.65 nm per scan. The resulting steps between the device and the TE line, BE line and field are 65, 55 and 105 nm, against 51, 56 and 103 nm measured by AFM: the topography is restored in sign and approximate size, but one rate cannot represent layers that sputter at different rates.

### C. Step 3: per-layer depth calibration

If layer thicknesses are known, each layer can be calibrated separately. Here we use the nominal thicknesses of the stack (Section IV-A and *Figure 3*): 60 nm top Pt, 30 nm $TaO_x$/Ta, 40 nm bottom Pt and 10 nm Ti. Because the sample is laterally structured, each region has its own stack: Pt/$TaO_x$/Ta/Pt/Ti in the device, Pt/$TaO_x$/Ta on the TE line, $TaO_x$/Ta/Pt/Ti on the BE line and $TaO_x$/Ta in the field. In each region the interfaces are located on the depth profile summed over the region interior, where counts are ample (Figure 5d for the device; all regions in supplementary Figure S3b), and every column is offset by its own substrate shift. With $K_m$ the scan of the $m$-th interface ($K_0 = 0$ at the surface) and $t_m$ the thickness of layer $m$, the sputter rate of that layer, in nm per cycle, is

$$r_m = \frac{t_m}{(K_m - K_{m-1})}, \tag{4}$$

and a voxel in layer $m$ ($K_{m-1} \leq k < K_m$) is placed at depth

$$d(x, y, k) = \sum_{l=1}^{m-1} t_l + (k - K_{m-1}) r_m, \tag{5}$$

Below the last interface the rate of the last layer is used. The corrected height of a voxel thus combines four elements: the shift of its column [Eq. (3)], the calibration of its region [Eqs. (4) and (5)], a substrate surface common to all regions, and one global offset that places the device top at its initial AFM height, because Tier 1 has no absolute height reference.

The calibrated rates expose two limits. First, the rates are only as good as the thicknesses: with 40 nm assumed for the bottom Pt, the device column gets 0.86 nm per cycle for that layer, whereas the AFM heights of the device top give a local rate of 0.58 nm per cycle between the stages at 157 and 195 cycles (Figure 6e), compared with the pooled Pt rate of 0.66 ± 0.06 nm per cycle fitted to all regions and stages (Section VIII-B). Second, and more serious, nothing constrains the rate below the last calibrated interface. In the field the 30 nm film is followed by the substrate, which sputters about twice as fast, so the field surface ends up 136 nm too high at the end of the profile. The initial steps, by contrast, come out close to the AFM values (50, 60 and 110 nm, against 51, 56 and 103 nm). The agreement of the initial steps supports the relative stack heights, but does not independently determine the thickness of each constituent layer or how depth is distributed within the profile; the test that does is comparison with the intermediate AFM surfaces, which the correction never sees (Section IX).

### D. Limits

Tier 1 also breaks down at the sidewalls, where the substrate signal is weak, the interface position changes abruptly and binned maps blur the transition. Columns within ≈0.6 μm of a region boundary are therefore excluded from the region statistics below. Tier 1 also requires a well-defined, flat substrate, so it does not apply to every sample. If the sample itself contains the element used to follow the substrate, the interface becomes hard to identify, and roughness or inhomogeneity of the substrate translates directly into errors in the corrected volume. If the sidewalls or the substrate sputter rate matter, a correlative tier is needed.

## VII. TIER 2 IN PRACTICE: SINGLE-STAGE CORRELATIVE CORRECTION

The correlative correction replaces the flat-substrate assumption with measured topography. It needs the surface before and after the profile on an absolute height scale (Section V-C) and the registration of that topography to the SIMS volume.

### A. Co-registration

AFM and SIMS images differ in field size, pixel size and possibly orientation. Both show the device, as a raised square in AFM and as the last column to reach the substrate in SIMS, so registration uses its outline. The four device edges are located with sub-pixel precision in the AFM image and in a SIMS map in which the device is distinct ($^{30}Si^{+}$ summed over scans 140–205), and a scale and an offset are fitted separately for each axis. Separate scales absorb calibration differences between the instruments, and the fitted scales, 0.496 and 0.495 SIMS pixels per AFM pixel, show that these are small here: the two agree within 0.2 % and lie within 1 % of the nominal 0.500, indicating close relative agreement between the AFM and SIMS coordinate scales; the different outline sizes along the two axes in Section V-A reflect the device outline, not the raster. Rotation is not fitted, since none above 0.4° was found. The mapping has as many parameters as fitted edges, so it has no fit residual; as a consistency check, after registration to the first stage the device edges of the other nine AFM images scatter by 0.05 µm RMS (at most 0.13 µm). Orientation is a setting. Correlating outline images for all eight combinations of transposition and mirror flips confirms that no transposition is needed (correlation 0.61–0.63, against 0.57–0.58 with transposition), but a symmetric device cannot reveal mirror flips; a wrong flip would affect only sub-micrometer features at the sidewalls. Each AFM image is smoothed to the SIMS pixel size before it is sampled onto the SIMS grid, to avoid aliasing.

### B. Surfaces before and after the profile

Each AFM image, registered to the first stage and leveled on the device top, gives heights relative to the device top, and the profiles give the absolute height of the device top at each stage (Section V-C). The surfaces before and after the profile are therefore

$$\begin{aligned} h_{before}(i,j) &= AFM_{before}(i,j) \\ h_{after}(i,j) &= AFM_{after}(i,j) - H_{sp} \end{aligned} \tag{6}$$

where $AFM_{before}$ and $AFM_{after}$ are the registered images placed on the absolute scale with the device-top height before sputtering (107.6 nm), and $H_{sp}$ = 145.7 nm is the sputtered depth of the device top measured on the profiles.

### C. Outside the AFM footprint

The 10 μm AFM image covers only the center of the 20 μm SIMS field; up to 6 μm around the device lies outside it, and there the surface has to be extrapolated from the areas that the AFM did measure. The extrapolation relies on the lateral uniformity of the sample away from the device: each region (TE line, BE line and field) has a single stack, so its surface outside the AFM image can be taken from the same region inside it. The SIMS data define the regions: the device as the last column to reach the substrate, the TE line by the absence of Ta at the start of the profile (its Pt covers the $TaO_x$/Ta film), and the BE line by its Ti (supplementary Figure S4).

The one complication is the trench next to the device (Section V-C): within a few micrometers of the device edge the surroundings are sputtered faster, so a region is not uniform there. The trench depth as a function of the distance from the device edge is measured on the long line profile, which runs along the TE line, and is assumed to be the same next to every region. A region is therefore extrapolated in two steps. Its height is read in the AFM image, in the strip of that region next to the device, and corrected for the trench depth at the distance of the strip; the trench depth at the distance of each extrapolated pixel is then added back. For the TE line, the long profile also gives the height away from the device directly, and that value is used instead; as a check, the TE-line height extrapolated from the image strip agrees with it within 1–5 nm at all stages. Measured and extrapolated surfaces are blended over four pixels at the border of the AFM image.

### D. Voxel remapping

Let the registered SIMS volume have $N_x \times N_y \times N_k$ voxels over a field $X \times Y$. Scan $k$ samples the surface left after $k$ sputter cycles, so voxel ($i$, $j$, $k$) is placed at

$$\begin{aligned} x &= \frac{iX}{N_x} \\ y &= \frac{jY}{N_y} \\ h(i,j,k) &= h_{before}(i,j) + \left(h_{after}(i,j) - h_{before}(i,j)\right) \times \frac{k}{N_k - 1} \end{aligned} \tag{7}$$

with $N_k - 1 = 239$ cycles between the two surfaces. Equation (7) assumes that the sputter front moves linearly in each column from the initial to the final surface. It corrects the initial topography and the mean sputter rate of each column, but not changes of rate within a column.

The intermediate AFM surfaces (Figure *6*), which Tier 2 does not use, test this assumption directly (supplementary Figure S5). On the device the predicted surface stays within 10 nm of the measured one at all stages. On the TE and BE lines next to the device, however, Tier 2 places the surface up to 35 and 40 nm too low (a surface-position error, at cycle 135): the lines sputter slowly through the $TaO_x$/Ta film and quickly through the substrate below it, while Eq. (7) spreads the total depth evenly over the whole profile.

## VIII. TIER 3 IN PRACTICE: MULTI-STAGE CORRELATIVE CORRECTION

### A. Multi-stage remapping

Tier 3 relaxes the linear assumption by measuring the surface during the SIMS measurement.[17] With surfaces $S_j(x, y)$ measured after $c_0 = 0 < c_1 < \ldots < c_J$ sputter cycles ($c_J = 239$ for the final surface; stage labels as in Section IV-C), each built over the full SIMS field as in Section VII-C, the height of voxel ($x$, $y$, $k$) is interpolated between the two neighboring surfaces:

$$h(x,y,k) = S_j(x,y) + \frac{k - c_j}{c_{j+1} - c_j}\left[S_{j+1}(x,y) - S_j(x,y)\right],\ \ c_j \le k \le c_{j+1} \tag{8}$$

With $J = 1$ this is Eq. (7); with intermediate surfaces the linear assumption is needed only within each segment. Here $J = 9$: eight intermediate stages at or near layer interfaces and the final surface.

### B. Sputter rates of the materials

Surface evolution depends on the local material-removal rate. The AFM images and the long profiles give the heights of all regions at every stage (*Figure 6a–d*). Fitting them with one rate per material, with the layer sequence of each region taken from the SIMS interfaces (Section VI-C), gives 0.36 ± 0.08 nm per cycle for $TaO_x$/Ta, 0.66 ± 0.06 nm per cycle for Pt and 0.90 ± 0.06 nm per cycle for the $SiO_2$ substrate (*Figure 6f*). The rates are fitted by unweighted least squares to the measured heights themselves, 40 values (four regions at ten stages), with one free starting height per region and one rate per material, so that each measurement enters the fit once. Each uncertainty is the larger of the residual-based standard error and a jackknife estimate obtained by leaving out one stage at a time; the jackknife also captures misfit that is correlated in time, such as the slow interval discussed below. These rates are pooled over all regions and stages; local rates scatter around them, for example 0.58 nm per cycle for the bottom Pt of the device (Section VI-C). The 10 nm Ti layer is too thin to be resolved between AFM stages (0.55 ± 0.25 nm per cycle). With these four rates the model reproduces the measured heights of all regions over the whole profile within 6.5 nm RMS (*Figure 6e*). The local rate maps show the same contrast directly: in each interval the regions sputtering $TaO_x$/Ta are the slowest and those sputtering $SiO_2$ the fastest (supplementary Figure S6).

These rate differences are also why Tier 2 fails next to the device (Section VII-D): the electrode lines sputter slowly through the $TaO_x$/Ta film and quickly through the substrate below it, so their surfaces descend unevenly while Tier 2 assumes a uniform descent, and the steps between the device and its surroundings change by 40–50 nm during the profile (*Figure 6e*).

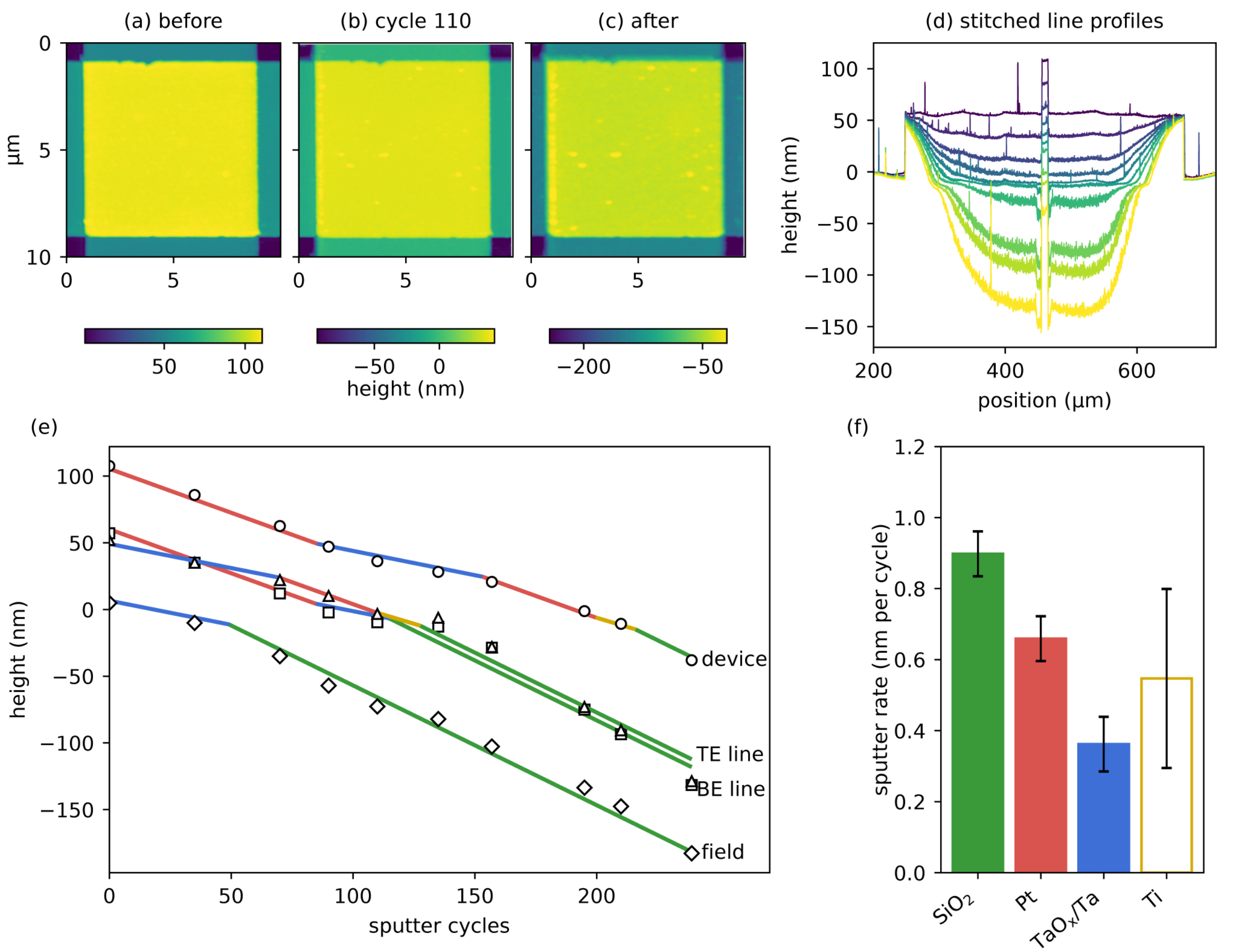

Figure 6. Multi-stage AFM series and sputter rates of the materials. As measured: (a)–(c) AFM topography before sputtering, after 110 cycles and after the profile, each on its own height scale (absolute heights, Section V-C); (d) stitched line profiles across the crater at all ten stages (dark to light). Derived: (e) heights of the device top and of the TE line, BE line and field away from the device (symbols, AFM) and the model from the material rates in (f) and the layer sequence of each region (lines, colored by the material being sputtered); (f) material sputter rates fitted to all stages and regions, with the uncertainties described in Section VIII-B (open bar: Ti, weakly constrained).

The model is only an approximation. Its largest misfit lies between cycles 110 and 135, when all regions sputter two to five times more slowly than before and after, whatever material they are in, for reasons we could not establish. The $SiO_2$ rate also differs between regions, from about 0.8 nm per cycle in the field to 1.2 nm per cycle on the lines, although the field heights, read in small corners of the AFM image, are the least certain. A correction based on constant material rates, or on two surfaces as in Tier 2, cannot follow such changes; Tier 3 uses the measured surfaces directly and needs neither assumption.

### Validation

Two tests use information that Tier 3 does not. The first is a comparison with the device model (Figure 3): interfaces that the model fixes across the device edge must come out right in the corrected volume. In the Tier 3 volume the $TaO_x$/Ta–bottom Pt and bottom Pt–Ti interfaces continue from the BE line into the device within 1 nm and the Ti–substrate interface within 9 nm; the top-Pt/$TaO_x$ interface lies 50 nm higher on the device than on the TE line, exactly the nominal thickness of the BE stack that the TE line climbs over; and the substrate interface lies at the same height in all four regions within 9 nm, against 28 nm after Tier 2 (Section IX). The second is a comparison with the intermediate AFM surfaces: each intermediate stage is left out in turn and predicted from the other nine (supplementary Figure S5). The left-out stage contributes nothing to its own prediction: every surface is built only from its own AFM image and long profile (its offset, trench and line height), and the quantities shared by all stages (the AFM-to-SIMS mapping, the registration reference and the profile datum) come from the first stage, which is never left out. Tier 2 likewise uses only the surfaces before and after the profile. Leave-one-out prediction of intermediate AFM surfaces gave RMS errors of 2.1–3.3 nm in the device and electrode-line interiors. This metric assesses surface interpolation; it does not establish the accuracy of individual layer thicknesses or sidewall geometry. The larger values in the field (14 nm RMS) come from the field corners inside the AFM image, which are only ≈1 μm wide and lie entirely in the trench zone. The corrected device column yields layer thicknesses of 57, 29, 26 and 12 nm for top Pt, $TaO_x$/Ta, bottom Pt and Ti, compared with nominal values of 60, 30, 40 and 10 nm. The present measurements do not establish the origin of the bottom-Pt discrepancy.

These results demonstrate that multi-stage interpolation accurately reproduces the measured surface evolution in the device and electrode-line interiors. However, the reported RMS errors characterize the prediction of region-median surface heights rather than the overall accuracy of the reconstructed 3D chemical volume. The latter additionally depends on the identification of material interfaces in the SIMS data, lateral resolution, and sputtering-induced changes in sample geometry.

### C. Practical limits

Each interruption costs time and adds a transient, a jump and a registration step (Section V-A). The partially sputtered surface must also survive repeated contact imaging. This device did, at ≈600 nN and without visible damage, but softer or poorly adhering layers may not, and intermittent-contact imaging or a sacrificial device should be tried first. Stages are best placed where the sputter rate of any region changes, i.e., at the interfaces seen in the SIMS profile; between stages the linear assumption still applies. Finally, Tier 3 remaps voxels vertically only, so lateral displacement at sloped sidewalls remains, and the sidewall band is still the least reliable part of the corrected volume.

## IX. COMPARISON OF THE CORRECTIONS

Figure *7* shows the four correction levels side by side on the same data, together with the idealized model of the device (flat layers with the nominal thicknesses of Figure *3* and vertical sidewalls). Without correction (*Figure 7b, e, j*), the volume is a flat slab: the stacks next to the device appear at the height of the device stack, and the substrate is stepped. Tier 1 (*Figure 7f,k*) levels the substrate and restores the device step, but misplaces the stacks next to the device

wherever the assumed sputter rates are wrong. Tier 2 (*Figure 7g,l*) ends on the correct final surface but pushes the stacks next to the device down, below the substrate of the device column. Finally, Tier 3 (Figure 7c, h and m) reproduces the structure of the model: the BE-line stack and its Ti continue from the line into the device, and the substrate interface is level across all regions.

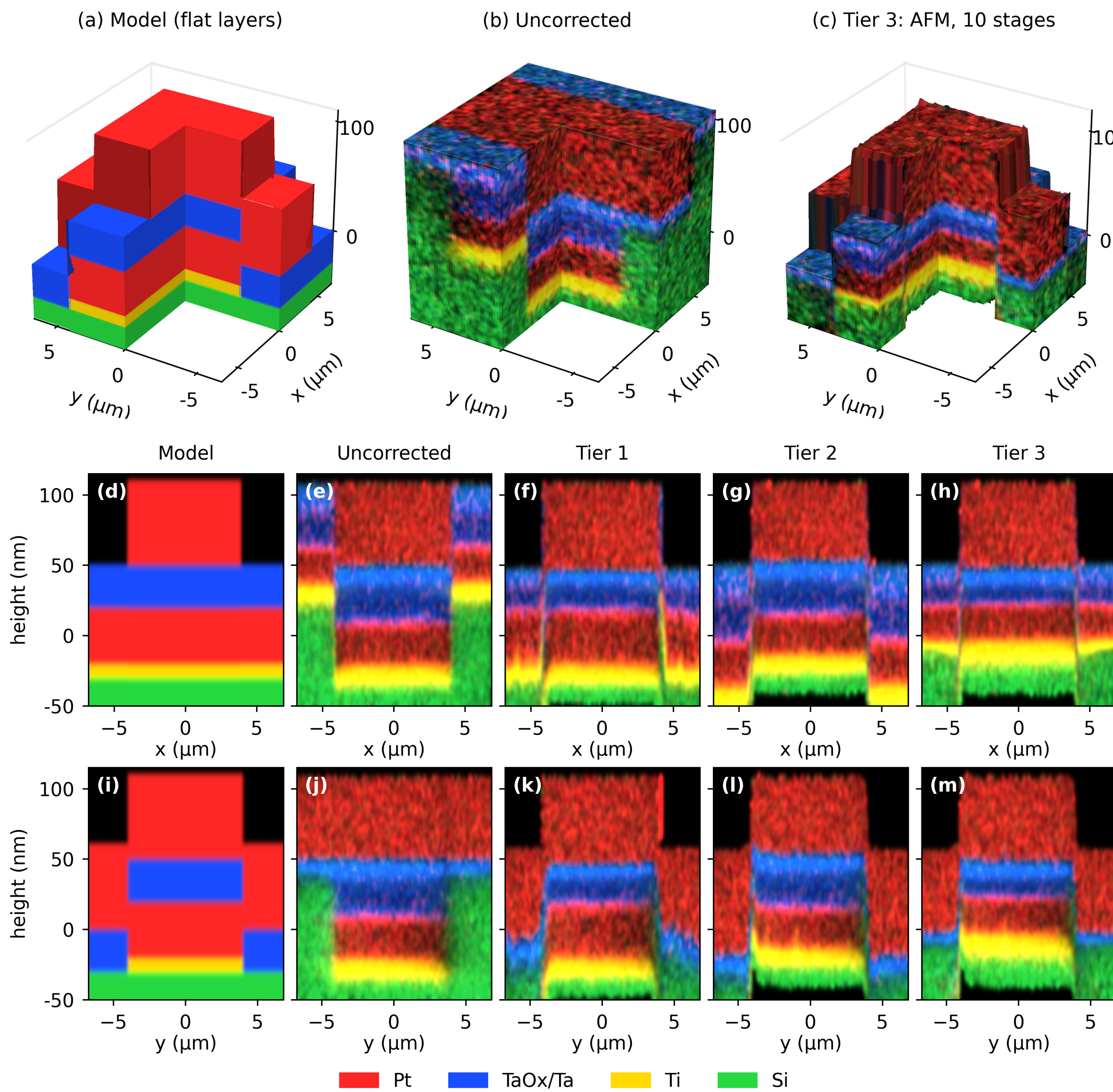

Figure 7. Topographic correction of the memristor device. (a)–(c) Cut-away 3D renderings of (a) the idealized model (flat layers, nominal thicknesses), (b) the uncorrected ToF-SIMS volume and (c) the volume after multi-stage correlative correction (Tier 3); heights exaggerated. (d)–(m) Cross sections through the device center along x (d)–(h) and y (i)–(m): (d), (i) model; (e),

(j) uncorrected; (f), (k) Tier 1 with per-layer calibration; (g), (l) Tier 2; (h), (m) Tier 3. Pt and $TaO_x$/Ta are unmixed abundances (Section V-B); Ti and Si are $^{48}Ti^+$ and $^{30}Si^+$. Displayed values are the mean intensity per voxel, with one brightness scale per layer for all data panels. Each voxel is taken to fill the slab between the surface sampled by its scan and the next one; values are resampled onto 1.5 nm height bins as thickness-weighted means (intensities, not counts), smoothed over 1.2 bins, with negative unmixing estimates clipped to zero and bins without data shown black. Sections are averaged over 1.3 µm and truncated at −50 nm; the substrate extends to the crater floor. The uncorrected and Tier 1 volumes are aligned to the initial AFM height of the device top.

Table *2* quantifies the comparison with two tests, and each catches something the other misses. The model test asks whether the corrected volume has the interface relationships expected from the nominal device geometry: the BE stack continuous from the line into the device, the TE stack raised by the BE stack it climbs over, and a level substrate. It needs only the design of the sample, so most readers can apply it. It separates the uncorrected volume (BE-stack jumps of ≈60 nm and a 105 nm substrate spread), Tier 2 (jumps of 25 nm, a TE step of 69 instead of 50 nm and a 28 nm substrate spread) and Tier 3 (≤9 nm, 50 nm and 9 nm). Substrate flatness cannot test either Tier 1 variant, because both impose it by construction. The TE step and the BE-stack continuity remain informative for Tier 1a, whereas Tier 1a+ imposes these relationships through the nominal layer thicknesses. The intermediate AFM surfaces show what this hides. Measured against them, Tier 1 reduces the error on the electrode lines about threefold relative to no correction, but is off by 46–58 nm RMS in the field, because nothing in the SIMS data constrains the sputter rate of the substrate below the 30 nm film. Tier 2 is right where it is anchored, at the start and end of the profile, and wrong in between, with surface-position errors of up to 40 nm on the lines, while Tier 3 stays within 2–3 nm RMS on the device and lines. In the field, where the AFM data are least reliable, Tier 3 (14 nm RMS) does not improve on Tier 2 (11 nm). On the device top, which is laterally uniform over 8 µm, every level, even the uncorrected volume, is within ≈11 nm RMS; the methods differ most strongly where the layer sequence changes laterally.

The field entries inside the AFM footprint come from corners only ≈1 µm wide, within the trench zone and next to two sidewalls, and are the least reliable values in Table *2*. Which level is enough depends on the question. For the qualitative layer sequence in the device interior, Tier 1 would suffice here; for the lateral relation between the device and its electrodes, for example to follow diffusion along the BE line, a correlative correction is needed, and Tier 3 when the layers sputter at very different rates.

Table 2. Comparison of the corrections on the memristor device. Model test (whole field): TE step, height of the top-Pt/$TaO_x$ interface on the device above that on the TE line (model 50 nm); BE-stack jump, largest jump of the three BE-stack interfaces between the BE line and the device (model 0); substrate spread, range of the region-median substrate heights (model 0). Intermediate AFM (inside the AFM image): RMS over the eight intermediate stages of the region-median difference between predicted and measured surface; Tier 3 is tested leave-one-out. All values for region interiors, ≥0.6 µm from boundaries. —, imposed by construction; n/a, not applicable.

| Correction | Information used | Model test (nm): TE step / BE-stack jump / substrate spread | Intermediate AFM, RMS error (nm): device / TE line / BE line / field |
|---|---|---|---|
| Model (expected) | nominal device geometry | 50 / 0 / 0 | n/a |
| Uncorrected (Tier 0) | SIMS; one sputter rate | −1 / 59 / 105 | 11 / 56 / 51 / 119 |
| Tier 1a: substrate-referenced | + nominal total thickness | 64 / 3 / — | 11 / 16 / 16 / 46 |
| Tier 1a+: per-layer calibration | + nominal layer thicknesses | 50 / 0 / — | 9.2 / 17 / 16 / 58 |
| Tier 2: correlative, single stage | AFM before and after | 69 / 25 / 28 | 6.5 / 21 / 27 / 11 |
| Tier 3: correlative, multi-stage | AFM at ten stages | 50 / 9 / 9 | 2.1 / 3.0 / 3.3 / 14 |

## X. PRACTICAL RECOMMENDATIONS

Table *3* collects the practical points from the worked example, in the order in which they arise. The central recommendation is independent validation: test a correction against information it did not use. A corrected volume can look convincing and still be wrong, and recovering the inputs of a correction does not test it.

Table 3. Checklist for topography-corrected 3D SIMS.

| Stage | Recommendation |
|---|---|
| Planning | Choose the correction tier before acquisition (Figure 2). Correlative tiers need topography at the same location before, during (Tier 3) and after sputtering; it cannot be added afterward. Place intermediate AFM stages where the sputter rate of any region changes. |
| Field size | Where drift and stage motion are expected, make the SIMS field larger than the AFM field by at least that margin, within the limits set by sampling and by how far the surroundings can be extrapolated. Use measured topography inside the AFM footprint and model the laterally uniform surroundings outside it. |
| Acquisition | Acquire enough scans per layer that interfaces are not limited by scan spacing. Note the mass resolution; at unit resolution, expect isobaric overlaps. Record SE or TIC images if a data-only morphology correction may be needed. |
| Registration | Register the image stack before correcting. Flag, but keep, transient scans after interruptions; model jumps and drift separately; use outline (gradient) images; apply integer shifts only, and report any resampling and its effect on counts. |
| Peak assignment | Check isotope ratios in layers of known composition. Remove overlaps by linear unmixing with reference patterns from regions of known composition, and verify the result on regions not used as references. |
| Interface detection | Bin and smooth only until the interface map is stable. At low counts or for gradual interfaces, use a threshold (half-rise) criterion instead of the steepest rise. Inspect the map, not only profiles. |
| Co-registration | Fit scale and offset per axis on sharp features such as device edges; check the orientation; report the residual misregistration. |
| Interpolation | Avoid cubic-spline upsampling across sharp steps; nearest-neighbor or linear interpolation does not overshoot. |
| Validation | Test against information the correction did not use: intermediate AFM surfaces (leave-one-out for Tier 3), flatness of a substrate known to be flat, independent crater depths. |

## XI. CONCLUSIONS

Topography and differential sputtering distort 3D SIMS reconstructions in predictable ways, and corrections exist at several levels of effort. We compared substrate-referenced reconstruction, with and without nominal layer-thickness calibration, with single-stage and multi-stage AFM correction on one dataset. Three preparation steps come first and apply to every tier: registration of the image stack, which in a profile interrupted for AFM means separating transients, jumps and drift; removal of isobaric overlaps, which at unit mass resolution can change the apparent layer sequence; and an absolute height scale for the AFM data. Data-only correction (Tier 1) needs no additional instrumentation and restores the topography of laterally uniform regions, but its depth scale is only as good as the assumed thicknesses and is unconstrained below the last known interface. Single-stage correlative correction (Tier 2) fixes the start and end surfaces but assumes uniform removal in between, which next to the device misplaces the predicted surfaces by up to 40 nm. Multi-stage correction (Tier 3) gives leave-one-out RMS surface-prediction errors of 2.1–3.3 nm in the device and electrode-line interiors, with a larger error of 14 nm in the small field corners, at the cost of instrument time and repeated imaging of a partially sputtered surface. Even then the corrected bottom Pt is 26 nm thick against 40 nm nominal, so an accurate surface does not guarantee accurate layer thicknesses. Which tier is needed depends on what the analysis requires and what the sample allows. Whatever the choice, a correction should be validated against information it did not use and reported with its inputs and residual errors. The data and code of this article are provided so that every step can be reproduced and adapted.

## SUPPLEMENTARY MATERIAL

See the supplementary material at [URL will be inserted by AIP Publishing] for the mass spectrum (Fig. S1), the reference patterns and checks of the unmixing (Table S1), the trench next to the device (Fig. S2), the effect of binning and smoothing on the Tier 1 interface map and the depth profiles of all regions used for the per-layer calibration (Fig. S3), the region masks and AFM footprint (Fig. S4), the predicted minus measured surfaces of all tiers at intermediate AFM stages (Fig. S5), local sputter-rate maps between AFM stages (Fig. S6), and the AFM topography before and after the profile (Fig. S7).

## ACKNOWLEDGMENTS

This work was conducted at the Center for Nanophase Materials Sciences, which is a DOE Office of Science User Facility, using instrumentation within ORNL's Materials Characterization Core provided by UT-Battelle, LLC, under Contract No. DE-AC05-00OR22725 with the U.S. Department of Energy.

## AUTHOR DECLARATIONS

*Conflict of Interest*

The authors have no conflicts to disclose.

*Author Contributions*

**Jacob M. Shusterman:** Investigation (lead); Formal analysis (equal); Writing – original draft (equal).
**Anton V. Ievlev:** Conceptualization (lead); Formal analysis (equal); Supervision (lead); Writing – original draft (equal); Writing – review & editing (lead).

## DATA AVAILABILITY

The data (SIMS counts before and after registration, AFM images and line profiles at all ten stages) and the Python code that reproduce every figure and number in this article are openly available in Zenodo at https://doi.org/10.5281/zenodo.23348079.[33]

# Supplementary Material

## Correcting topographic artifacts in three-dimensional ToF-SIMS: A practical guide from data-only reconstruction to correlative AFM

Jacob M. Shusterman and Anton V. Ievlev[a)]
*Center for Nanophase Materials Sciences, Oak Ridge National Laboratory, 1 Bethel Valley Road, Oak Ridge, Tennessee 37831*
[a)]Email: ievlevav@ornl.gov

This document contains one table and seven figures for the worked example. All were produced by the Python code provided with the article (scripts/run_example.py and scripts/fig_si.py), which is openly available together with the data at https://doi.org/10.5281/zenodo.23348079; section numbers refer to the main text.

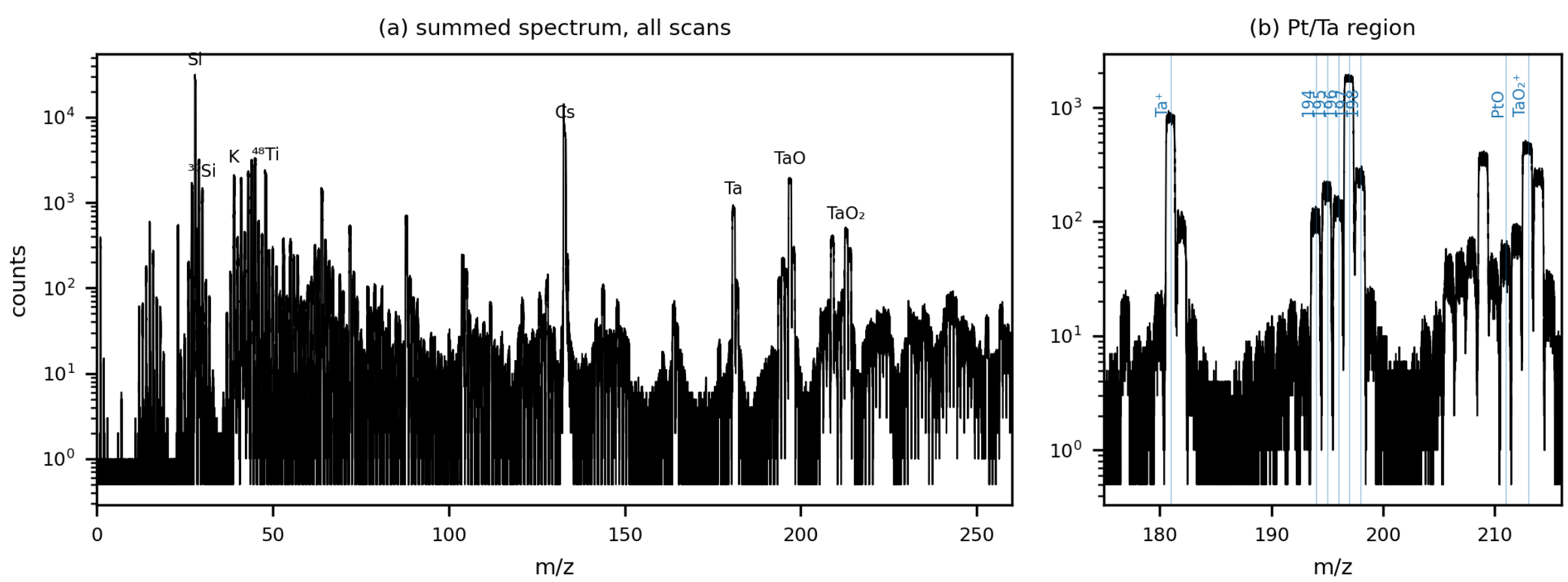

FIG. S1. Positive-ion mass spectrum summed over all pixels and scans: (a) 0–260 u and (b) the Pt/Ta region. Unbunched acquisition gives flat-topped peaks at unit mass resolution, so $Pt^+$ isotopes (*m*/*z* 194–196 and 198), Ta adduct ions, $TaO^+$, $^{196}PtH^+$ and $TaOH^+$ overlap (Section V-B). $Cs^+$ (*m*/*z* 133) and $K^+$ (*m*/*z* 39) are contaminants.

TABLE S1. Linear unmixing (Section V-B). Top: reference patterns, as the fraction of counts of each material in each mass channel, measured in regions of known composition (Pt: top Pt of the device, scans 15–70; $TaO_x$: field film, scans 6–30; Ta: Ta layer of the device, scans 118–142; region interiors). Bottom: normalized signal fractions after unmixing (signal estimates, not compositions) in check regions that were not used to define the patterns.

| Channel (m/z) | Pt (%) | TaOx (%) | Ta (%) |
|---|---|---|---|
| Pt194 | 27.9 | 0.7 | 0.8 |
| Pt195 | 30.9 | 2.9 | 3.5 |
| Pt196 | 24.0 | 1.9 | 1.9 |
| TaO | 4.4 | 52.8 | 42.2 |
| TaOH | 6.8 | 7.3 | 5.4 |

| Channel (m/z) | Pt (%) | TaOx (%) | Ta (%) |
|---|---|---|---|
| Ta | 1.1 | 20.0 | 43.4 |
| TaO2 | 1.6 | 13.3 | 2.3 |
| PtO195 | 3.3 | 1.1 | 0.4 |
| Check region | Pt (%) | TaOx (%) | Ta (%) |
| device TaOx/Ta | -0.4 | 65.7 | 34.7 |
| device bottom Pt | 100.5 | -1.3 | 0.8 |
| TE line Pt | 101.3 | 0.1 | -1.5 |
| BE line TaOx | 0.6 | 94.9 | 4.5 |

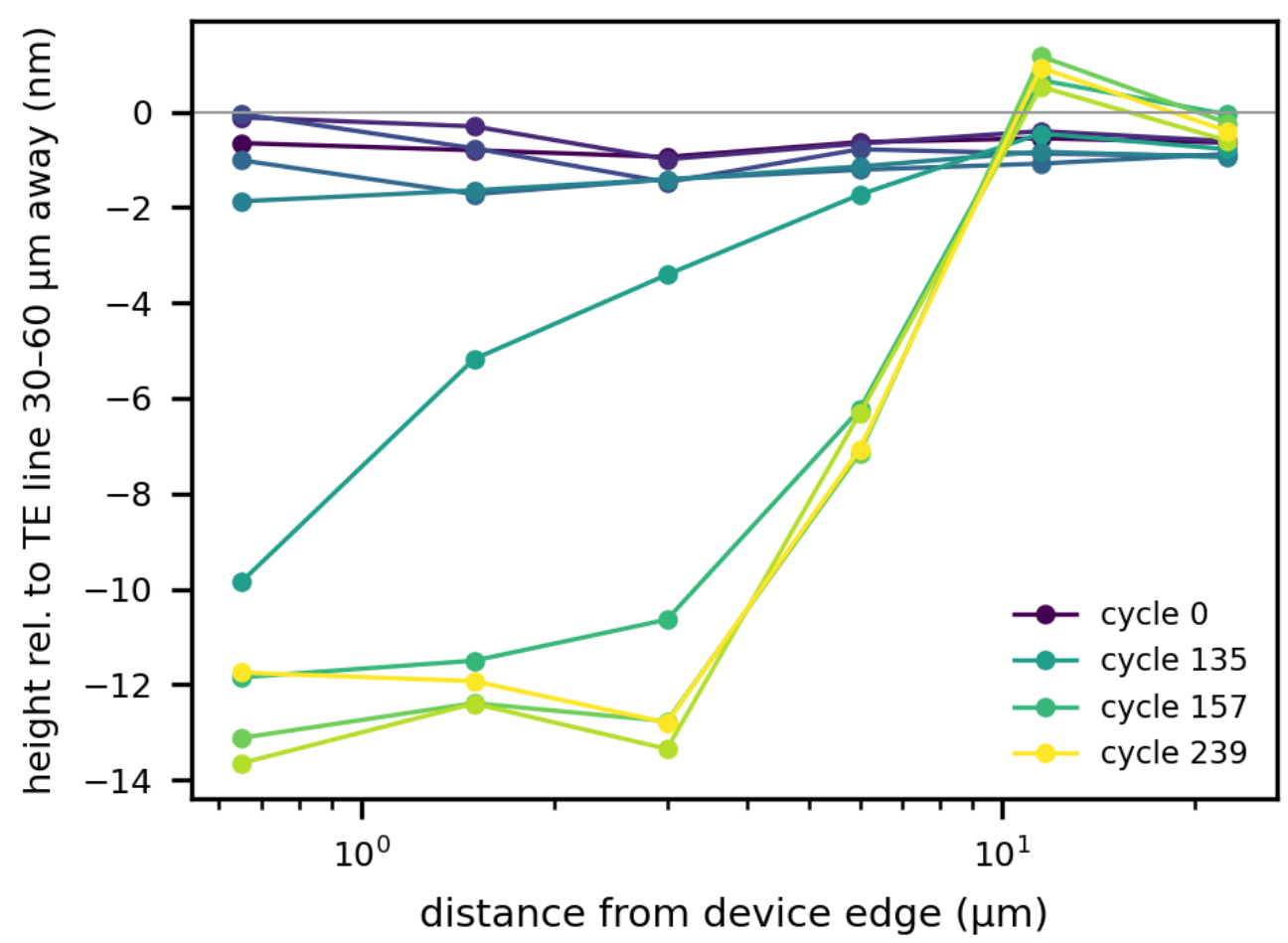

FIG. S2. Height of the TE line next to the device relative to the same line 30–60 µm away, as a function of the distance from the device edge, at all ten AFM stages (long line profiles, both sides of the device pooled). A trench of up to ≈20 nm forms from the Ta stage (cycle 135) onward and extends to ≈8 µm.

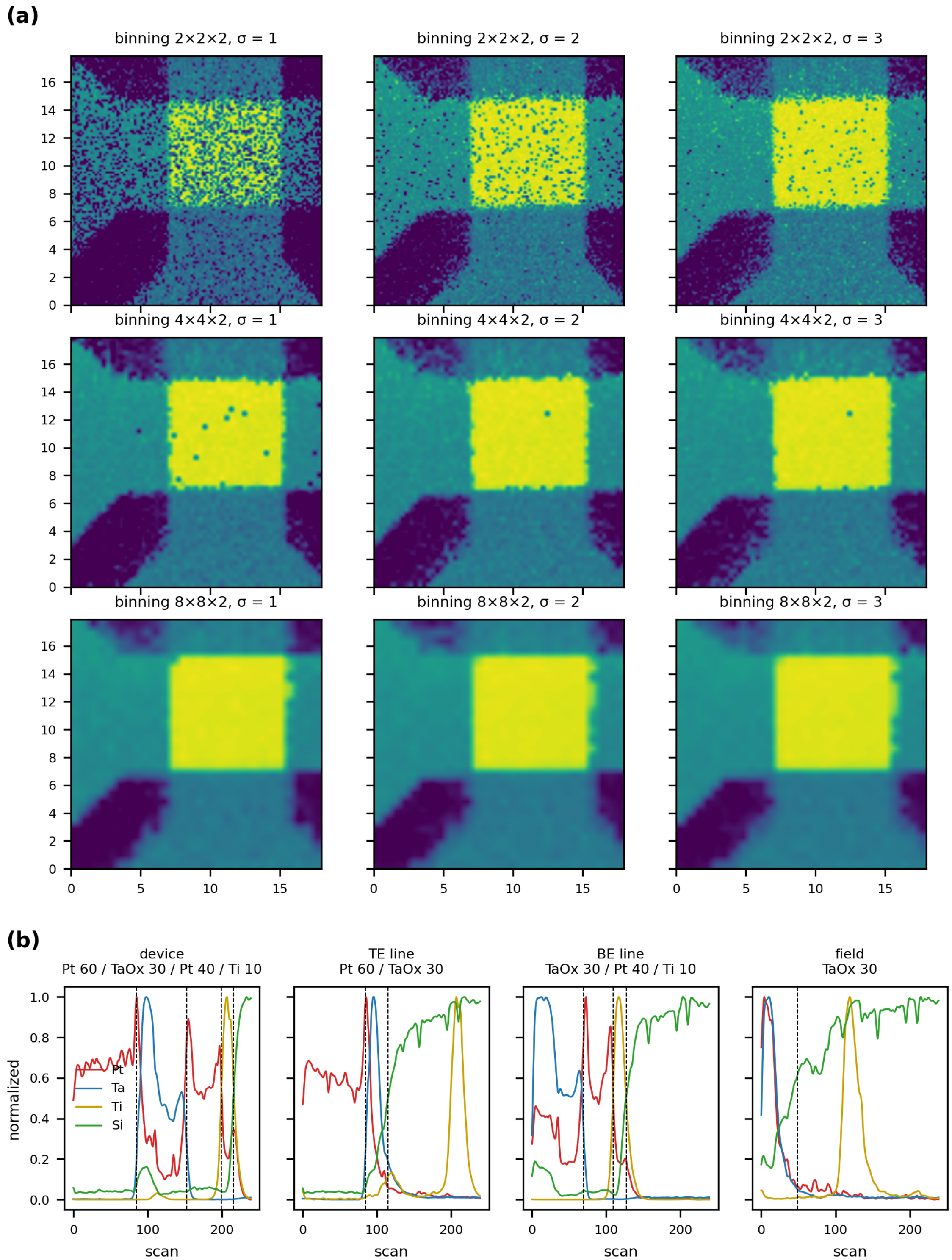

FIG. S3. Tier 1 details. (a) Substrate-interface maps $K_s$ (half-rise criterion, $^{30}Si^+$) for binning 2 × 2 × 2, 4 × 4 × 2 and 8 × 8 × 2 (rows) and Gaussian smoothing over σ = 1, 2 and 3 bins (columns); common color scale, 40–225 scans. The main text uses 4 × 4 × 2 and σ = 2: weaker binning or smoothing leaves noise, stronger blurs the device edges. (b) Normalized depth profiles summed

over the interior of each region, with the interfaces used for the per-layer calibration (dashed) and the nominal stacks (titles, nm). The $^{48}Ti^{+}$ peaks deep in the TE-line and field profiles lie in the substrate.

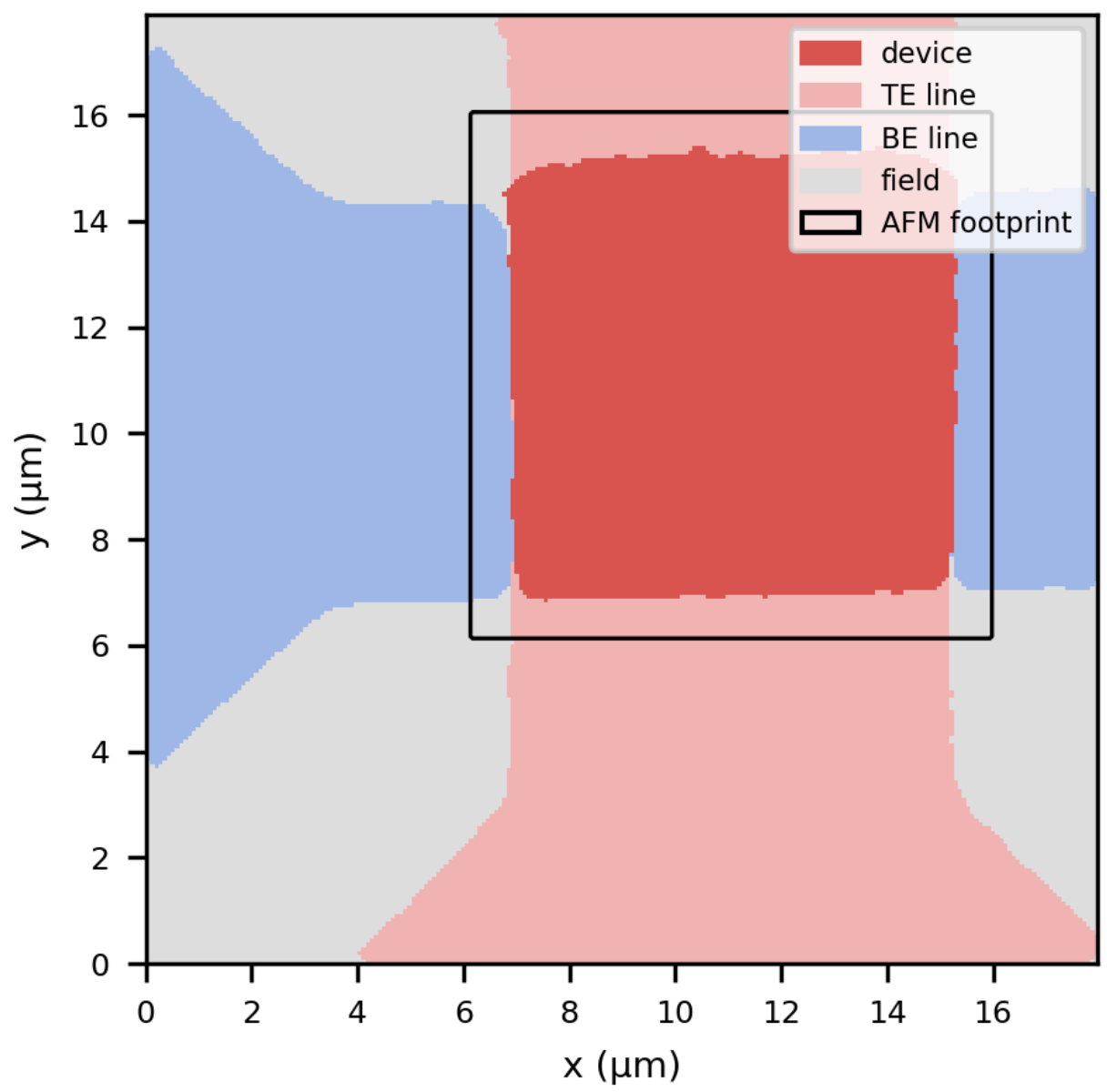

FIG. S4. Region masks derived from the SIMS data (device: last column to reach the substrate; TE line: no Ta at the start of the profile; BE line: Ti) and the area covered by the AFM image (black outline) on the registered SIMS grid (Section VII-C).

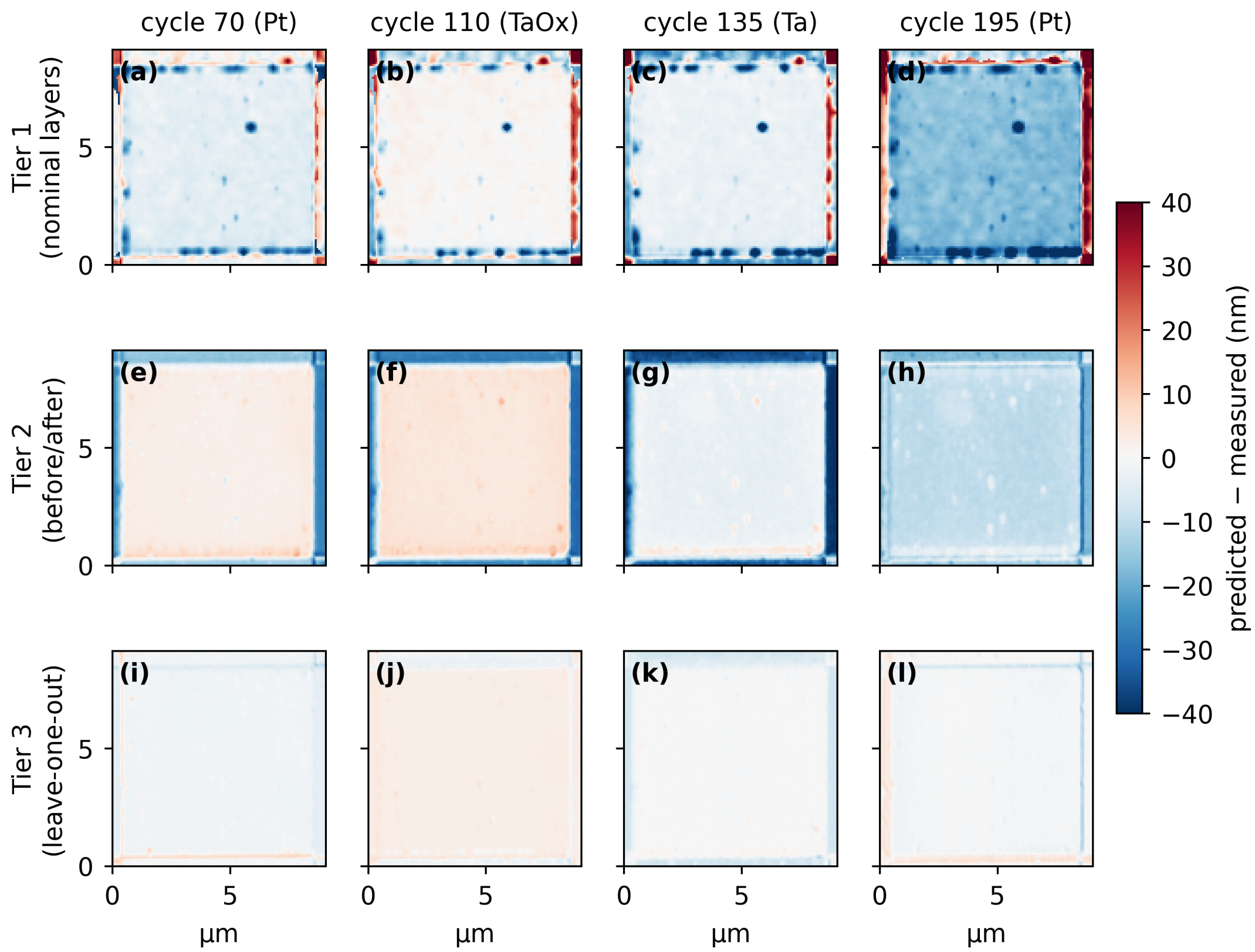

FIG. S5. Predicted minus measured surface at four intermediate AFM stages, inside the AFM image: (a)–(d) Tier 1 with per-layer calibration (nominal thicknesses); (e)–(h) Tier 2; (i)–(l) Tier 3, each stage predicted from the other nine (leave-one-out). Region statistics are given in Table 2 of the main text.

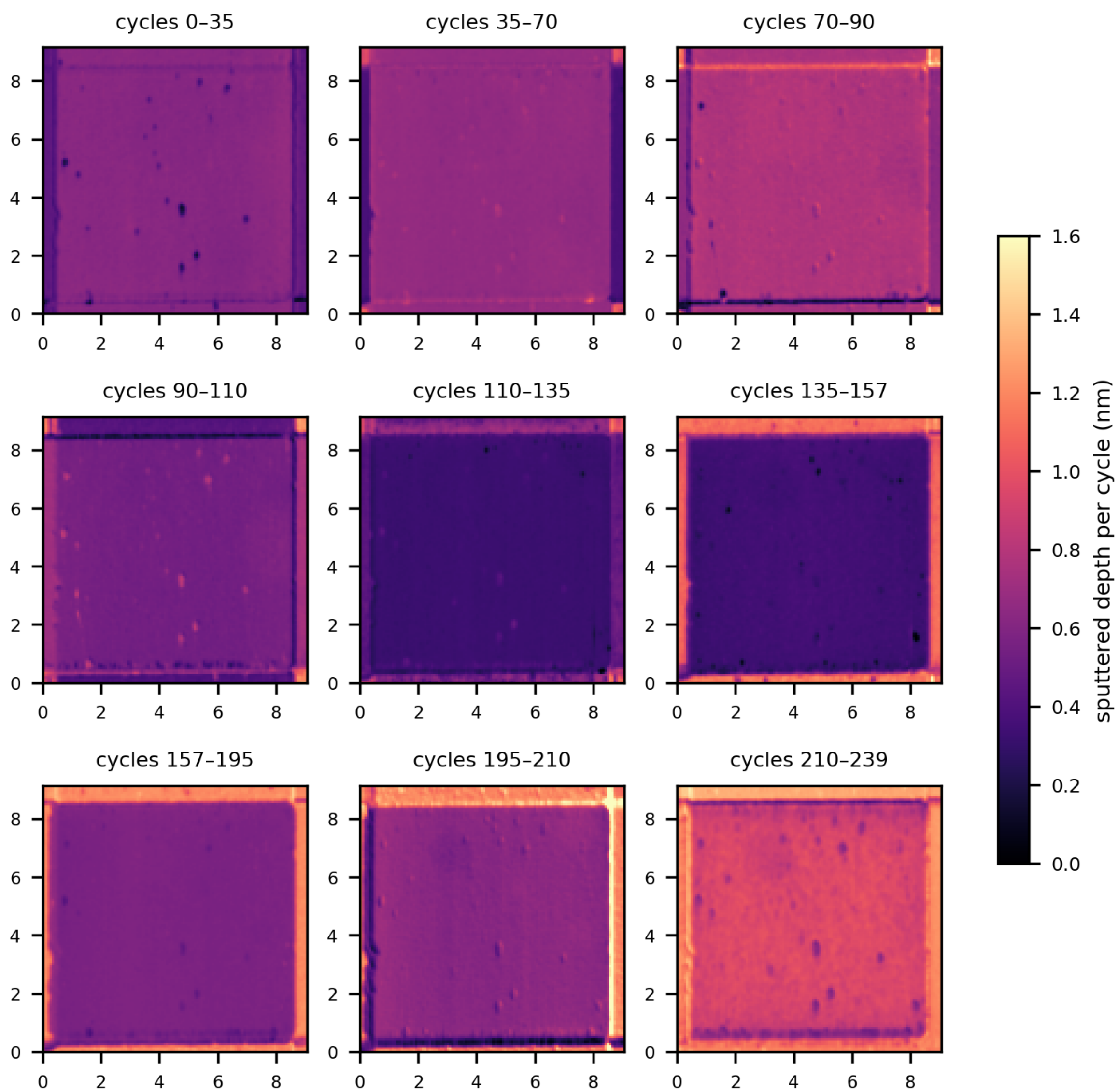

FIG. S6. Local sputter-rate maps: sputtered depth per cycle between consecutive AFM stages, inside the AFM image, computed from the full-field surfaces of Section VII-C; the material rates fitted to all regions and stages are in Figure 6f of the main text. Bright and dark lines along the device edges combine residual sub-pixel misregistration and the trench next to the device.

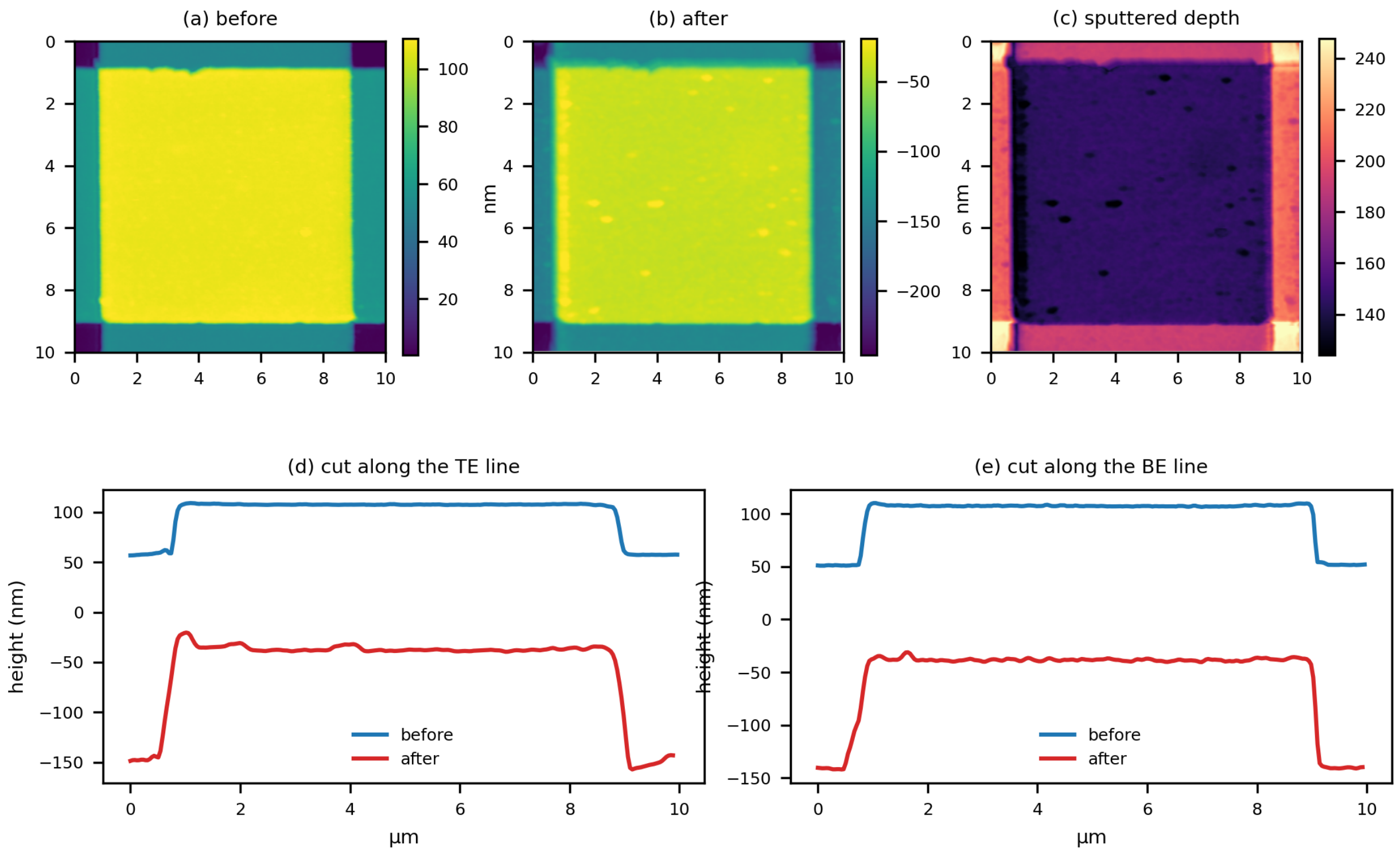

FIG. S7. AFM topography (a) before and (b) after the profile on the absolute height scale (Section V-C), (c) sputtered depth, and cuts through the device center (d) along the TE line and (e) along the BE line. By the end of the profile one sidewall has broadened to ≈0.8 μm and carries a small rim, while the opposite sidewall stays steep, consistent with oblique ion incidence.